\documentclass[12pt]{article}
\usepackage{a4wide}
\usepackage{epsfig,graphicx}
\usepackage{amsmath}
\usepackage{amsfonts}
\usepackage{amssymb}
\usepackage{array}
\usepackage{slashed}
\usepackage{pstricks}
\usepackage{multirow}
\usepackage{float}
\usepackage{fancyhdr}
\usepackage[latin1]{inputenc}
\usepackage[english]{babel}

\newcommand{\beq}{\begin{equation}}
\newcommand{\eeq}{\end{equation}}
\newcommand{\bea}{\begin{eqnarray}}
\newcommand{\eea}{\end{eqnarray}}

\newcommand{\OMIT}[1]{{}}

\newcommand\spur{\raise.15ex\hbox{/}\kern-.57em }

\newcommand{\lsim}{
\mathrel{\hbox{\rlap{\hbox{\lower4pt\hbox{$\sim$}}}\hbox{$<$}}}}
\newcommand{\gsim}{
\mathrel{\hbox{\rlap{\hbox{\lower4pt\hbox{$\sim$}}}\hbox{$>$}}}}

\begin{document}

\begin{flushright}
IFIC-09-14\\
FTUV-09-0402\\
KEK-TH-1308 \\
LA-UR-09-01473\\
TU-845
\end{flushright}
\vspace{0.5 true cm}
\begin{center}
{\Large {\bf On the model discriminating power}}\\
\vspace{0.5 true cm}
{\Large {\bf 
of $\mu \to e$ conversion in nuclei
}}\\

\vspace{1.5 true cm}
{\large Vincenzo Cirigliano$^{a}$, 
Ryuichiro Kitano$^{a,b}$, \\
Yasuhiro Okada$^{c}$, 
Paula Tuzon$^{a,d}$}\\
\vspace{1.5 true cm}
${}^a$   
{\sl Theoretical Division, Los Alamos National Laboratory, Los Alamos, NM 87545, USA}
\vspace{0.5 true cm}

${}^b$   
{\sl Department of Physics, Tohoku University, Sendai 980-8578, Japan}
\vspace{0.5 true cm}

${}^c$ {\sl  Theory Group, KEK, Oho 1-1, Tsukuba, Ibaraki 305-0801, Japan } {\em and}  \\
{\sl  Department of Particle and Nuclear Physics, The Graduate University for Advanced Studies, 
 Oho 1-1, Tsukuba, Ibaraki 305-0801, Japan} \\
\vspace{0.5 true cm}

${}^d$ {\sl  Departament de F\'{\i}sica Te\`orica, IFIC, 
Universitat de Val\`encia -- CSIC  \\ 
Apt.  Correus  22085, E-46071 
Val\`encia, Spain} \\ 
\vspace{0.5 true cm}
\end{center}
\vspace{.5cm}

\begin{abstract}
We assess the model discriminating power of 
a combined phenomenological analysis of 
$\mu \to e \gamma$ and  $\mu \to e$ conversion on different target nuclei, 
including the current hadronic uncertainties. 
We find that the theoretical uncertainties can be largely reduced by 
using input from lattice QCD and do not constitute a limiting factor  in 
discriminating models where one or at most two underlying operators 
(dipole, scalar, vector)  provide the dominant source of lepton flavor violation. 
Our results show that  a realistic discrimination among underlying mechanisms 
requires a measurement of  the ratio of conversion rates 
at the $5 \%$ level (two light nuclei)  or at the  $20 \%$ level (one light and one heavy nucleus). 
We have illustrated these main conclusions 
also  in the context of a supersymmetric  model.

\end{abstract}

\thispagestyle{empty}
\newpage


\newpage
\setcounter{page}{1}

\section{Introduction}

Lepton Flavor Violating (LFV) decays of charged leptons  provide 
a theoretically clean probe of physics  beyond the Standard Model (SM),   
due to the un-observably small  branching fractions ($\sim 10^{-50}$) within 
the SM, minimally extended to include massive neutrinos.  
Searches for SM  forbidden muon processes,  such as $\mu \to e \gamma$,  
$\mu \to e \bar{e} e$, and $\mu \to e$ conversion in nuclei, 
have provided so far  the strongest constraints  on  LFV dynamics, 
with 90$\%$ C.L. upper limits given by 
$B_{\mu\to e\gamma}< 1.2\times 10^{-11}$~\cite{muegamma99}, 
$B_{\mu^+ \rightarrow e^+e^-e^+} < 1.0 \times 10^{-12}$~\cite{mu3e88},
$B_{\mu\to e} ({\rm Au}) < 8 \times 10^{-13}$~\cite{mueconvAu}, 
$B_{\mu\to e}({\rm Ti}) <4.3 \times 10^{-12}$ ~\cite{mueconvTi}, 
$B_{\mu\to e} ({\rm Pb}) < 4.6 \times 10^{-11}$ ~\cite{mueconvPb}~\footnote{
$B_{\mu \to e} (Z,A)$ represents the ratio of $\mu \to e$ conversion rate  over 
 muon capture rate, namely $ \frac{ \Gamma_{conv}(Z,A) }{\Gamma_{capt} (Z,A) }$.}.

It is a well known fact that while the decay $\mu \to e \gamma$ is 
only sensitive to  a transition magnetic dipole operator, both 
$\mu \to e \bar{e} e$ and $\mu \to e$ conversion in nuclei 
are sensitive to transition charge radii operators as well as 
purely contact four-fermion interactions induced by physics beyond the 
SM.    In other words,  different LFV decays 
have different sensitivities to underlying LFV mechanisms (effective 
operators).  
This leads naturally to ask the question whether one could 
infer the  relative strength of these different  operators 
in a completely phenomenological and model-independent way. 
This would  allow one to discriminate among different underlying models of LFV 
and thus would provide valuable input for model building. 

In Ref.~\cite{ryuichiro} it was  pointed out that in principle, 
by combining  the rates of $\mu \to e \gamma$ and $\mu \to e$ conversion 
on different target nuclei, one could obtain information on underlying
models.
There are three types of effective operators that contribute to
      the coherent $\mu \to e$ conversion process: the dipole, the
      vector, and the scalar operators. In the non-relativistic
      approximation of the muon wave function, the three operators give
      the same form of overlapping integrals among the wave functions of
      the initial muon and the final electron and the nucleon density in
      the target nuclei. However, as the relativistic and finite nuclear size effects become
      important for heavy nuclei~\cite{ryuichiro,shanker,marciano},  
      the transition amplitudes for the three
      operators show different dependences on the atomic number $Z$.
      The relative numbers of neutrons and protons also change as $Z$
      increases. This fact helps to find out if the
      lepton-flavor-violating operators couples to up-type or down-type
      quarks again by looking at the target atom dependence.
In this work we go back to this issue with the aim to 
\begin{itemize}
\item  quantify the theoretical  uncertainty 
induced by the quark scalar density matrix elements in the nucleon;  
\item  quantify the experimental 
precision required to realistically infer useful information on the underlying 
LFV mechanisms. 
\end{itemize}

We organize our discussion as follows:  in Section~\ref{sect:leff}   
we review the derivation of the $\mu \to e$ conversion rate 
starting from a general effective theory description of the LFV physics. 
In Section \ref{sect:test1}   we explore the phenomenological consequence of 
the simplest possible models, in which only one effective LFV operator dominates. 
We extend this analysis in Section \ref{sect:test2}  to the class of models in which two
operators dominate. 
In Section \ref{sect:susy} we specialize our discussion  to a  supersymmetric (SUSY)  model and 
summarize the conclusions of our analysis in Section \ref{sect:conclusions}.

\section{LFV effective interaction and the $\mu \to e$  conversion rate}
\label{sect:leff}

In this section we review the procedure to calculate the rate 
of the $\mu \to e$ conversion in nuclei,  starting from a general parameterization of
new physics effects via effective operators at a scale $\Lambda $ larger 
than the electroweak scale $v \simeq 174$ GeV.
 
\subsection{Effective Lagrangian}
We start with the most general effective Lagrangian which describes LFV
transitions between charged leptons of first and second families at the
weak scale:
\begin{eqnarray}
\mathcal L_{eff}^{(q)} & = & - \frac{1}{\Lambda^2} \ \Bigg[ \left( C_{DR} m_{\mu} \; \bar{e}\sigma^{\rho\nu} P_L \mu + C_{DL} m_{\mu} \; \bar{e}\sigma^{\rho\nu}  P_R \mu \right) F_{\rho\nu} 
\nonumber\\
 &+&
\sum_{q} \left( C^{(q)}_{VR}  \; \bar{e}\mathcal \gamma^{\rho}P_R \mu  +  C^{(q)}_{VL}  \; \bar{e}\mathcal \gamma^{\rho} P_L \mu \right) \bar{q}\gamma_{\rho}q 
\nonumber\\ 
&+& 
\sum_{q} \left( C^{(q)}_{SR} m_{\mu} m_qG_F \; \bar{e} P_L \mu + C^{(q)}_{SL} m_{\mu} m_qG_F \; \bar{e} P_R \mu \right) \bar{q}q 
\nonumber\\ 
&+&
\left( C_{GR} m_{\mu}G_F \; \bar{e} P_L \mu + C_{GL} m_{\mu}G_F \; \bar{e} P_R \mu \right) \frac{\beta_H}{2g_{s}^3} G^{\rho\nu}_aG^a_{\rho\nu} +h.c.  ~ \Bigg] ~.
\label{eq:weakscaleL}
\end{eqnarray}
We have not included operators involving $\bar{q} \gamma_\rho \gamma_5 q$, 
$\bar{q}  \gamma_5 q$,  or  $\bar{q} \sigma_{\rho \nu}  q$ since they do not contribute 
to the coherent conversion processes.
In the above expression $\Lambda$ represents the scale where new
physics effects appear. We take $\Lambda \equiv 1$~TeV in this paper.
The $C_{AB}$'s are dimensionless constants containing information about
the underlying theory; the subindexes $R,L$ correspond to the chirality
of the final electron which is determined by $P_{R,L}=(1\pm
\gamma^5) / 2$, $q$ are light and heavy quarks.
The field strength of the photon and the gluon are defined by
$F_{\rho\nu}=\partial_{\rho}A_{\nu}-\partial_{\nu}A_{\rho}$ and
$G_{\rho\nu}^a=\partial_{\rho}G_{\nu}^a-\partial_{\nu}G_{\rho}^a
-f_{abc}G_{\rho}^bG_{\nu}^c$,
respectively. The normalization is chosen so that the kinetic terms are
given by $-(1/4)FF$ and $-(1/4g_s^2)G^a G^a$.
The $\sigma$ matrix is defined by
$\sigma^{\rho\nu}=\frac{i}{2}[\gamma^{\rho},\gamma^{\nu}]$. 
$G_F = 1/(2 \sqrt 2 v^2)$  is the Fermi constant, while
$m_\mu$ and $m_q$ represent the muon and running quark masses at $\mu = \Lambda$,
respectively. 
We have introduced  the running quark masses and the beta function of the QCD
coupling constant, $\beta = (g_s^3/16 \pi^2) (11 - 2N_F/3)$, so that the
coefficients $C$'s do not depend on the renormalization scale under 
QCD running at 1-loop level.
The notation $\beta_{H,L}$ is used to distinguish the Lagrangian with
all quarks contributions (H) from the one where heavy quarks are
integrated out (L).  The Lagrangian in Eq.~\ref{eq:weakscaleL}
describes three kind of
interactions that violate the lepton flavor: The effective interaction
with a photon (Dipole term), the effective interaction with quarks
(Scalar and Vector terms) and the effective interaction with gluons
(Gluon term).  

In order to evaluate the $\mu \to e$ conversion rate, it is appropriate
to use the effective Lagrangian at the nucleon level~\cite{Kosmas:2001mv}. 
We first integrate
out the heavy quarks before matching to the nucleon level Lagrangian. It
can be straightforwardly done by using the matching of the trace
anomaly~\cite{Shifman:1978zn}. The Lagrangian is given by
\begin{eqnarray}\label{qlagrangian}
\mathcal L_{\rm eff}^{(q')}&=&- \frac{1}{\Lambda^2} \Bigg[ 
\left( C_{DR} m_{\mu} \; \bar{e}\sigma^{\rho\nu} P_L \mu + C_{DL} m_{\mu} \; \bar{e}\sigma^{\rho\nu} P_R \mu \right) F_{\rho\nu} 
\nonumber\\ 
&+& 
 \sum_{q=u,d,s} \left( C^{(q)}_{VR}  \; \bar{e}\mathcal \gamma^{\rho}P_R \mu  + 
 C^{(q)}_{VL}  \; \bar{e}\mathcal \gamma^{\rho} P_L \mu \right) \bar{q}\gamma_{\rho}q 
 \nonumber\\ 
 &+& 
 \sum_{q=u,d,s} \left( C^{(q)}_{SR} m_{\mu} m_qG_F \; \bar{e} P_L \mu + C^{(q)}_{SL} m_{\mu} m_qG_F \; \bar{e} P_R \mu \right) \bar{q}q 
 \nonumber\\ &+&
 \left( C_{GQR} m_{\mu}G_F \; \bar{e} P_L \mu + C_{GQL} m_{\mu}G_F \; \bar{e} P_R \mu \right) \frac{\beta_L}{2g_{s}^3} G^{\rho\nu}_aG^a_{\rho\nu} +h.c.~ \Bigg],
\end{eqnarray}
where $\beta_L$ is the beta function of  three-flavor QCD. The new
coefficients of the gluon terms are expressed in terms of the original
Lagrangian parameters as follows:
\begin{eqnarray}
C_{GQR} &=& \sum_{Q=c,b,t} C_{SR}^{(Q)} \; \kappa_Q + C_{GR} \kappa \nonumber\\
C_{GQL} &=& \sum_{Q=c,b,t} C_{SL}^{(Q)} \; \kappa_Q + C_{GL} \kappa \; ,
\end{eqnarray}
where 
\begin{eqnarray}
 \kappa_Q = {\Delta (\beta / g_s^3) \over (\beta / g_s^3)_L}
 = {2 \over 27},\ \ \ 
  \kappa = {(\beta / g_s^3)_H \over (\beta / g_s^3)_L} = {7 \over 9}.
\end{eqnarray}

The Lagrangian \eqref{qlagrangian} can be evolved with the
renormalization group down to energy scales of the order of $\mu \sim 1$
GeV,  by simply taking the quark masses and gauge coupling constants 
in the Lagrangian to be the running ones at $\mu \sim 1$ GeV. 
 At this low scale, we match to the effective Lagrangian written in
terms of the relevant degrees of freedom, namely nucleons, leptons, and
photons.
That can be done by the following replacements of operators:
\begin{eqnarray}
m_q \; \bar{q}q  &\rightarrow& f_{SN}^{(q)} \; 
m_N \; {\bar{\psi}}_N {\psi}_N  
\nonumber\\  
\bar{q}\gamma_{\rho} q  &\rightarrow& f_{VN}^{(q)} \; 
{\bar{\psi}}_N \gamma_{\rho} {\psi}_N \; 
\nonumber\\ 
\frac{\beta_L}{{2g_{s}^3}} GG  &\rightarrow& f_{GN} \; m_N 
\; {\bar{\psi}}_N {\psi}_N \; .
\label{eq:replace}
\end{eqnarray} 
where $N$ represents each nucleon ($N=p,n$), $\psi_N$ are the nucleon
fields, and $f$'s are nucleon form factors. The form factors  depend in
principle on the momentum transfer, which we will neglect as it is
smaller than the typical scale of the nucleon structure.
The fact that $\langle N|\theta_{\alpha}^{\alpha}|N\rangle=m_N \;
\langle N | \bar{\psi}_N\psi_N | N \rangle$ ($\theta_{\alpha}^\alpha$ is the trace of the energy 
momentum tensor)  implies the simple sum-rule
\begin{eqnarray}
	1 = \sum_{q=u,d,s} f^{(q)}_{SN} + f_{GN} \; ~, 
\end{eqnarray}
which we use to eliminate the form-factor $f_{GN}$ in terms of the scalar nucleon form 
factors $f^{(q)}_{SN}$. 
The nucleon vector form factors are known from the vector current conservation,
\begin{eqnarray}
	\begin{array}{cc} f_{Vp}^{(u)}=2 & f_{Vn}^{(u)}=1 \\ f_{Vp}^{(d)}=1 & f_{Vn}^{(d)}=2 \\ f_{Vp}^{(s)}=0 & f_{Vn}^{(s)}=0 \end{array} \; ,
\end{eqnarray}
while the calculation of the scalar form factors $f^{(q)}_{SN}$ is
non-trivial.  As discussed below, in our analysis we will use input from
Chiral Perturbation Theory and the lattice QCD to asses the impact of
current and future uncertainties on the conversion rate. \\

Collecting the above results, the Lagrangian at nucleon level can be
written as
\begin{eqnarray}
\mathcal L_{eff}^{(N)}&=&-\frac{1}{\Lambda^2} \sum_{N=p,n} \Bigg[ \left( C_{DR} m_{\mu} \; 
\bar{e}\sigma^{\rho\nu} P_L \mu + C_{DL} m_{\mu} \; \bar{e}\sigma^{\rho\nu}
 P_R \mu \right) F_{\rho\nu} 
\nonumber\\
&+&
\left( \tilde{C}^{(N)}_{VR}  \; \bar{e} \gamma^{\rho} P_R \mu  + \tilde{C}^{(N)}_{VL}  
\; \bar{e} \gamma^{\rho}  P_L \mu \right)  \ \bar{\psi}_N \gamma_{\rho} 
\psi_N 
\nonumber\\ 
&+&
G_F  m_\mu  m_N  \, \left( \tilde{C}^{(N)}_{SR}  \; \bar{e} P_L \mu + 
  \tilde{C}^{(N)}_{SL}  \; \bar{e} P_R \mu \right) \; \bar{\psi}_N \psi_N 
+  h.c.\Bigg] \; .
\end{eqnarray} 
The new effective couplings $\tilde{C}$'s contain the information about
the underlying theory as well as the form factors. The vector
couplings are:
\begin{eqnarray}
\tilde{C}^{(p)}_{VR} &=& \sum_{q=u,d,s} C^{(q)}_{VR} \; f^{(q)}_{Vp} \\
\tilde{C}^{(n)}_{VR} &=& \sum_{q=u,d,s} C^{(q)}_{VR} \; f^{(q)}_{Vn} \\
\tilde{C}^{(p)}_{VL} &=& \sum_{q=u,d,s} C^{(q)}_{VL} \; f^{(q)}_{Vp} \\
\tilde{C}^{(n)}_{VL} &=&  \sum_{q=u,d,s} C^{(q)}_{VL} \; f^{(q)}_{Vn} ~, 
\end{eqnarray}
while the scalar ones read:
\begin{eqnarray}
\tilde{C}^{(p)}_{SR} &=&  \sum_{q=u,d,s} C^{(q)}_{SR} \; f^{(q)}_{Sp} 
\ + \ C_{GQR} \; (1-\sum_{q=u,d,s} f^{(q)}_{Sp}) \\
\tilde{C}^{(n)}_{SR} &=&  \sum_{q=u,d,s} C^{(q)}_{SR} \; f^{(q)}_{Sn} 
\ + \ C_{GQR} \; (1-\sum_{q=u,d,s} f^{(q)}_{Sn}) \\
\tilde{C}^{(p)}_{SL} &=& \sum_{q=u,d,s} C^{(q)}_{SL} \; f^{(q)}_{Sp} 
\ + \ C_{GQL} \; (1-\sum_{q=u,d,s} f^{(q)}_{Sp}) \\
\tilde{C}^{(n)}_{SL} &=&  \sum_{q=u,d,s} C^{(q)}_{SL} \; f^{(q)}_{Sn} 
\ +  \ C_{GQL} \; (1-\sum_{q=u,d,s} f^{(q)}_{Sn}) ~.
\end{eqnarray}

\subsection{Transition rates}
The nucleon-level effective Lagrangian can be used to take matrix
elements at the atomic and nuclear level. In the non-relativistic
approximation, the relevant matrix elements are
\begin{eqnarray}
	\langle A,Z|\bar{\psi}_p\psi_p|A,Z\rangle&=&Z\rho^{(p)} \nonumber\\
	\langle A,Z|\bar{\psi}_n\psi_n|A,Z\rangle&=&(A-Z)\rho^{(n)} \nonumber\\
	\langle A,Z|\bar{\psi}_p\gamma^0\psi_p|A,Z\rangle &=&Z\rho^{(p)} \nonumber\\
	\langle A,Z|\bar{\psi}_n\gamma^0\psi_n|A,Z\rangle &=&(A-Z)\rho^{(n)} \nonumber\\
	\langle A,Z|\bar{\psi}_N\gamma^i\psi_N|A, Z \rangle &=&0 \; .
\end{eqnarray}
Here $|A,Z\rangle$ represents the nuclear ground state, with $A$ and $Z$ the mass and atomic number of the isotope, while $\rho^{(p)}$ and $\rho^{(n)}$ are the proton and neutron densities respectively.
The conversion rate of the process is written as
\begin{eqnarray}
\Gamma_{conv} &=& \frac{m_{\mu}^{5}}{4\Lambda^4} \left| C_{DR} \; D + 
4 G_F m_\mu  \left( m_p    \tilde{C}^{(p)}_{SR}  S^{(p)} +  m_n  \tilde{C}^{(n)}_{SR}   \; S^{(n)} 
\right)
+   \tilde{C}^{(p)}_{VR} \; 4V^{(p)} + \tilde{C}^{(n)}_{VR} \; 4V^{(n)} \right|^2 
\nonumber \\
&+&  \frac{m_{\mu}^{5}}{4\Lambda^4} \left| C_{DL} \; D + 
4 G_F m_\mu  \left( m_p    \tilde{C}^{(p)}_{SL}  S^{(p)} +  m_n  \tilde{C}^{(n)}_{SL}   \; S^{(n)} 
\right)
+   \tilde{C}^{(p)}_{VL} \; 4V^{(p)} + \tilde{C}^{(n)}_{VL} \; 4V^{(n)} \right|^2 
\nonumber \\
\end{eqnarray}
in terms of the dimensionless integrals $D, V^{(N)}, S^{(N)}$, representing 
the overlap of electron and muon wavefunctions weighted by appropriate 
combinations of protons and neutron densities~\cite{ryuichiro}. 
For phenomenological applications, it is useful to normalize the
conversion rate to the muon capture rate, introducing the quantity:
\begin{equation}
B_{\mu \to e} (Z)  \equiv   \frac{ \Gamma_{conv}(Z,A) }{\Gamma_{capt} (Z,A) }~.
\end{equation}

Finally, we note here the branching ratio for the purely radiative process  
$\mu \to e \gamma$ in terms of the effective couplings defined above:
\begin{equation}
B_{\mu \to e \gamma} \equiv \frac{\Gamma (\mu \to e \gamma)}{ 
\Gamma (\mu \to e \nu_\mu \bar{\nu}_e)}   =  \frac{48  \pi^2}{ G_F^2  \, \Lambda^4} 
\ \left( |C_{DR}|^2   + |C_{DL}|^2 \right)  ~. 
\end{equation}

\subsection{Sources of uncertainty} 

There are two sources of uncertainty in the calculation of the
transition rate: (i) scalar form factors and (ii) neutron density (for
high $Z$ nuclei).  The latter uncertainty has been carefully discussed in
Ref.~\cite{ryuichiro}, where several approaches to determine the neutron
density have been reviewed and used in the calculation of the overlap
integrals.  Whenever data from polarized proton scattering exists, the
uncertainty on the overlap integrals $S^{(n)}$ and $V^{(n)}$ can be
reduced to a few percent even for heavy nuclei such as Pb.  Otherwise,
it should be considered to be of the order of $10 \%$.  In this work we
focus on the uncertainty induced by the scalar density matrix elements in the nucleon.

The scalar form factors defined in Eq.~(\ref{eq:replace}) can be
re-expressed in terms of ratio of quark masses and ratios of nucleon
matrix elements as follows~\cite{cors-nath}:
\begin{eqnarray}
f_{Sp}^{(u)} &=&   \frac{m_u}{m_u + m_d}   \, \left(1 + \xi \right) \, \frac{\sigma_{\pi N}}{m_p}  \\
f_{Sp}^{(d)} &=&   \frac{m_d}{m_u + m_d}   \, \left(1 - \xi \right) \, \frac{\sigma_{\pi N}}{m_p}  \\ 
f_{Sp}^{(s)} &=&   \frac{m_s}{m_u + m_d}   \, y  \, \frac{\sigma_{\pi N}}{m_p} \\
f_{Sn}^{(u)} &=&   \frac{m_u}{m_u + m_d}   \, \left(1 - \xi \right) \, \frac{\sigma_{\pi N}}{m_p}  \\
f_{Sn}^{(d)} &=&   \frac{m_d}{m_u + m_d}   \, \left(1 + \xi \right) \, \frac{\sigma_{\pi N}}{m_p}  \\ 
f_{Sn}^{(s)} &=&   \frac{m_s}{m_u + m_d}   \, y  \, \frac{\sigma_{\pi N}}{m_p} ~, 
\end{eqnarray}
where
\begin{eqnarray}
\sigma_{\pi N} &=&   \frac{m_u + m_d}{2} \, \langle p | \bar{u} u  + \bar{d} d | p\rangle \\ 
\xi &=& \frac{\langle p | \bar{u}u  - \bar{d} d | p \rangle }{ \langle p | \bar{u}u + \bar{d} d | p \rangle } \\
y &=&   \frac{2 \langle p | \bar{s} s | p \rangle }{ \langle p | \bar{u}u + \bar{d} d | p \rangle }~.
\label{eq:ypar}
\end{eqnarray}

Information on the above matrix elements can be obtained 
from  $\pi N$ scattering data,  from an 
analysis of the octet baryon masses within Heavy Baryon Chiral
Perturbation Theory, or from Lattice QCD.

For the $\sigma$-term, we will use the lattice result~\cite{onogi}
\beq
\sigma_{\pi N} = (53  \pm 2 ({\rm stat }) ^{+21} _{-7}  ({\rm syst})) \, {\rm MeV}~, 
\eeq
whose uncertainty covers determinations from $\pi N$ scattering~\cite{gls,pavan},
from ChPT analysis of baryon masses~\cite{borasoy}, 
as well as from previous lattice analyses~\cite{procura04}.   
For the ratio measuring isospin-breaking, we will use~\cite{cors-nath,cheng89}:
\beq
\xi = 0.132 \pm 0.035 ~.    
\eeq
For the ratio $y$ quantifying the strange quark content of the nucleon,  the situation is less clear. 
A Chiral Perturbation Theory analysis gives the range $y = 0.21 \pm 0.2$~\cite{borasoy}.
The large uncertainty reflects the poor knowledge of the relevant low-energy constants, 
even within resonance saturation (the matching renormalization scale is arbitrary). 
A recent lattice QCD analysis~\cite{onogi} of  the matrix element  
$\langle N | \bar{s} s | N \rangle$  
within the overlap fermion formulation with  two dynamical flavors
leads to  
\beq
y  =  0.030  \pm 0.016 ({\rm stat }) ^{+0.006} _{-0.008}  ({\rm extrap})
^{+0.001} _{-0.002}  (m_s)~.
\label{eq:ylatt}
\eeq
This result is  obtained from the lattice matrix element  $\langle N | \bar{s} s | N \rangle$  
by dividing out the sigma-term as calculated in the same lattice simulation. 
Therefore, the uncertainty in $f_{SN}^{(s)} \propto  y \times \sigma_{\pi N}$ 
is controlled by Eq.~\ref{eq:ylatt}, with $\sigma_{\pi N} = 53$ MeV simply  providing the normalization.
The lattice result is consistent with the Chiral Perturbation Theory range,  although 
suggesting a much smaller strange content of the nucleon.  The difference with respect to
previous lattice results has been attributed to a lattice artifact
(mixing with wrong chirality operator) in the Wilson fermion approach.
The uncertainty on this value is at the moment dominated by statistics. 
 
For the purpose of this work, we will vary the  parameter $y$  
within  both a  ``conservative'' range and an ``optimistic'' range. 
For the conservative range we take  $y \in [0,0.4]$, which coincides with the ChPT 
range of Ref~\cite{borasoy}. 
For the  optimistic range we take $y \in [0,0.05]$ which reflects more closely the recent 
JLQCD result~\cite{onogi}  and seems a realistic guess of the uncertainty that will be reached 
by lattice calculations in the next decade.  
 
Finally, for the ratios of quark masses, we  use the the input~\cite{leutwyler}
\bea
\frac{m_u}{m_d} &=& 0.553 \pm 0.043 \\
\frac{m_s}{m_d} &=&  18.9  \pm 0.8~.
\eea

\section{Testing the  single operator dominance hypothesis} 
\label{sect:test1}

We now turn to illustrate the model discriminating power of 
a combined phenomenological analysis of 
$\mu \to e \gamma$ and  $\mu \to e$ conversion on different target nuclei. 
In order to organize  the discussion, we define here four classes of models,  
in which only one underlying short distance  operator dominates over all the others.  
We call these four classes of models the ``single-operator'' 
dominance models. We will first analyze this 
simplest class of models and then consider the more involved case 
in which two operators have comparable strengths and interference effects cannot 
be neglected. 

\subsection{Dipole, Vector  and Scalar models}
\label{sect:models1}
  
\begin{itemize}
\item{\textbf{Dipole model}}

The Dipole model is defined by the assumption that, among all LFV 
short-distance operators,  the dipole operator is the dominant one. 
For simplicity, we focus on the case in which the outgoing lepton has definite 
chirality~\footnote{Allowing for the presence of  outgoing leptons with both chiralities 
(e.g.  both $C_{DR} \neq 0$  and $C_{DL} \neq 0$) 
would not change the conclusions of the single-operator analysis of this section.}. 
Explicitly, in terms of the effective couplings defined in Eq.~\ref{eq:weakscaleL}, 
this class of models is defined by: 
\begin{eqnarray}
	\begin{array}{cc} 
	C_D \equiv  C_{DR}  \neq 0 
	 & C_{else}=0 \end{array} \; .
\end{eqnarray}
Most supersymmetric scenarios, including SUSY-GUT models~\cite{GUT1} and 
SUSY see-saw models~\cite{seesaw1} fall in this class of models.

\item{\bf{Vector model 1: $V^{(\gamma)}$}}

This model is defined by the assumption that the transition charge radius  operator  gives the 
dominant contribution to the LFV lagrangian. 
The model is defined by
\begin{eqnarray}
	\begin{array}{lll} 
	C_V \equiv C_{VR}^{(u)} =  - 2 \,  C_{VR}^{(d)}    \neq 0,  &   \ \  \   &  C_{else}=0 \; ,\end{array} 
\end{eqnarray}
and is  explicitly realized in large regions of  the Left-Right symmetric model parameter 
space~\cite{Cirigliano:2004mv}. In this model $\tilde{C}_{VR}^{(p)} \neq 0$, while 
$\tilde{C}_{VR}^{(n)} = 0$.

\item{\bf{Vector model 2:  $V^{(Z)}$}}

The Vector model 2 is defined by the assumption that the  underlying dominant operator is 
an effective  $Z$-penguin.  The ratios of couplings of different quarks is governed by the  
couplings of the $Z^0$ coupling to quarks.
The model is defined by:
\begin{eqnarray}
	\begin{array}{lll} C_V \equiv C_{VR}^{(u)} 
 = \displaystyle\frac{C_{VR}^{(d)}}{a} \neq 0,  &  \ \ \  &  C_{else}=0 \; , \end{array} 
\end{eqnarray}
where $a$ is the ratio of the \emph{down} and \emph{up} quarks coupling to the $Z$-boson:
\begin{eqnarray}
a=
\frac{T^3_{d_L}+T^3_{d_R}-(Q_{d_L}+Q_{d_R})
\sin^2\theta_W}{T^3_{u_L}+T^3_{u_R}-(Q_{u_L}+Q_{u_R}) \sin^2\theta_W}  = -1.73  \; .
\end{eqnarray}
With this value of $a$ (corresponding to $\sin^2 \theta_W = 0.223$)  we obtain 
$\tilde{C}_{VR}^{(n)} / \tilde{C}_{VR}^{(n)}  = - 9.26$,  in contrast to the $V^{(\gamma)}$ model.

\item{\textbf{Scalar model}}

This model is defined by:
\begin{eqnarray}
	\begin{array}{ll} C_S \equiv C_{SR}^{(d)} = C_{SR}^{(s)} = C_{SR}^{(b)} \neq 0, & \ \  C_{else}=0 \; . \end{array} 
\end{eqnarray}
This model may be explicitly realized in some
regions of the usual R-parity conserving SUSY see-saw parameter
     space~\cite{largetanbeta1}
(large $\tan \beta$ and relatively low "heavy"  Higgs sector)  and  
within R-parity violating SUSY~\cite{RPVSUSY1}.
\end{itemize}

\begin{figure}[t]
\centering 
\includegraphics[width=7.5cm]{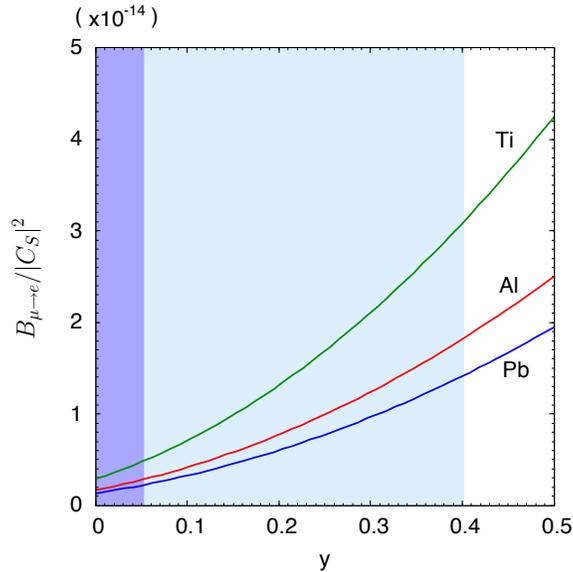}
\caption{The $y$-parameter dependence of the conversion branching ratio
in the scalar model.}
\label{fig:y-dependence}
\end{figure}

Among the above models, the scalar model suffers from the uncertainty in
the $y$ parameter. We show in Fig.~\ref{fig:y-dependence} the
$y$-parameter dependence of the conversion branching ratio. The
uncertainty is quite large if we take the conservative range, $y \in
[0,0.4]$.

Each of the above classes of models has only one free parameter -- the
ratio $C_{i}/\Lambda^2$ of the dominant effective coupling over the
square of the new physics scale.  It is clear, then, that the
single-operator dominance hypothesis makes parameter-free predictions
for ratios of LFV branching fractions and therefore it can be tested so
long as two LFV rates are measured.
We will discuss how well one can distinguish models in the presence of
the theoretical uncertainties.

\subsection{$\mu \to e \gamma$ vs $\mu \to e$ conversion} 

\begin{figure}[t]
\centering 
\includegraphics[width=10.5cm]{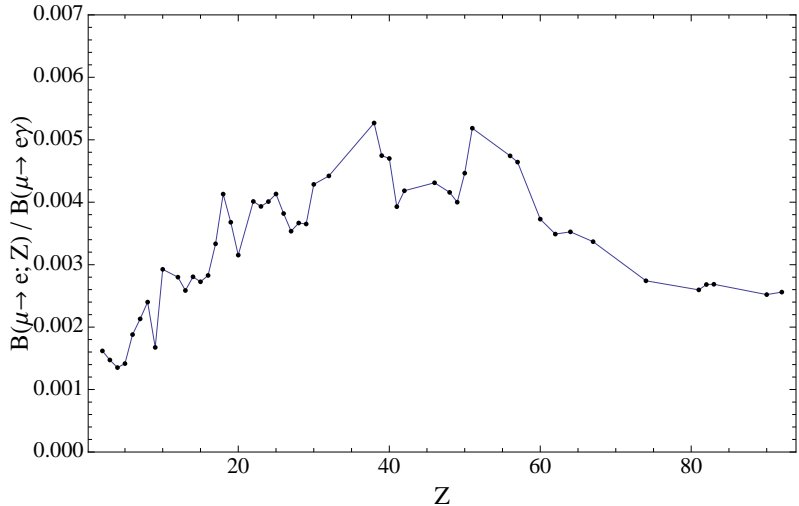}
\caption{Ratio $R(Z)$ of $\mu \to e$ conversion over $B(\mu \rightarrow e \gamma)$ 
versus $Z$ in the case of Dipole dominance model.}
\label{fig:ratiomeg}
\end{figure}

If $\mu \to e \gamma$ and $\mu \to e$ conversion in at least one target nucleus 
are observed,  this immediately  opens up  the possibility to test the Dipole dominance 
model.  In fact, in this model the ratio 
\beq
R (Z) = \frac{B_{\mu \to e} (Z)}{B_{\mu \to e \gamma}}   
\eeq
is entirely fixed by the overlap integrals $D$~\cite{ryuichiro}, which 
are essentially free of  theoretical uncertainty.
$R(Z)$ is predicted to scale as ${\cal O} (\alpha/\pi)$ and we plot it in Fig.~\ref{fig:ratiomeg}. 
We omit from the plot the points corresponding to $^{166}_{68} {\rm  Er}$, $^{181}_{73} {\rm Ta}$, 
and  $^{197}_{79} {\rm Au}$, as data on the nucleon densities are either obtained from 
quite old experiments or not well established~\cite{dejager}. 
Any deviation from the pattern shown in Fig.~\ref{fig:ratiomeg} would imply 
the presence of scalar or/and  vector contributions. 
In order to disentangle these possibilities, one needs to study the target 
dependence of the conversion rate. 

\subsection{Target dependence of $\mu \to e$ conversion} 

\begin{figure}[t]
\centering \includegraphics[width=10.5cm]{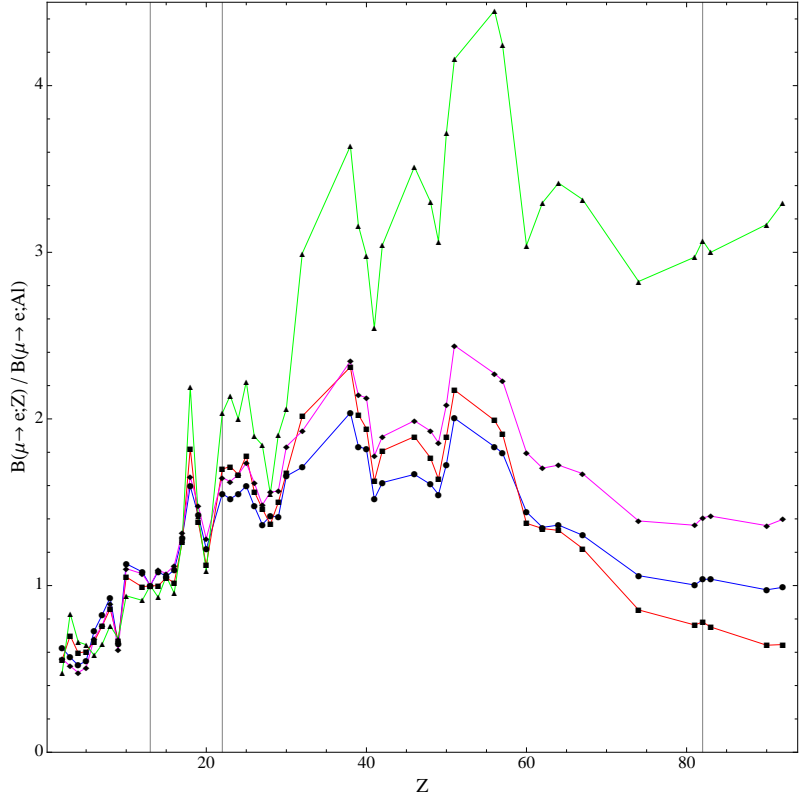} \caption{Target
dependence of the $\mu \to e$ conversion rate in different
single-operator dominance models.  We plot the conversion rates
normalized to the rate in Aluminum ($Z=13$) versus the atomic number $Z$  for
the four theoretical models described in the text: $D$ (blue), $S$
(red), $V^{(\gamma)}$ (magenta), $V^{(Z)}$ (green).  The vertical lines 
correspond to $Z=13 \,  ({\rm Al})$, 
$Z=22  \, ({\rm Ti})$, and 
$Z=83\,  ({\rm Pb})$.}
\label{fig:target-dependence-1}
\end{figure}

In principle, any single-operator model can be tested with two
conversion rates, even if $\mu \to e \gamma$ is not observed. To
illustrate this point, we update the analysis of Ref.~\cite{ryuichiro}
and plot in Fig.~\ref{fig:target-dependence-1} the conversion rate
(normalized to the rate in Aluminum) as a function of the $Z$ of the
target nucleus, for the four classes of single-operator models defined
above.  Compared to Ref.~\cite{ryuichiro}, the novelty here is the
inclusion of a second vector model ($V^{(Z)}$).

The results of Fig.~\ref{fig:target-dependence-1} show some noteworthy
features.  First, we note the quite different target dependence of the
conversion rate in the two vector models considered.  
This can be  understood as follows: in the case of the $V^{(\gamma)}$ model,
the behavior in Fig.~\ref{fig:target-dependence-1} simply traces the
$Z$-dependence of $V^{(p)}$ (the photon only couples to the protons in the
nucleus).  On the other hand, in the case of the $V^{(Z)}$ model, the $Z$
boson couples predominantly to the neutrons in the nucleus and the  
target dependence of the ratio $V^{(n)}/V^{(p)}\sim (A-Z)/Z$ generates the
behavior observed in Fig.~\ref{fig:target-dependence-1}.

Next, let us focus on the actual discriminating power of the
$Z$-dependence.  Clearly, the plot shows that the model-discriminating
power tends to increase with $Z$. This is a simple reflection of the fact
that the whole effect is of relativistic origin and increases in heavy
nuclei.  So in an ideal world, in order to maximize the chance to
discriminate among underlying models, one would like to measure the
conversion rate in a light nucleus, say Aluminum or Titanium, as well as
in a large-$Z$ nucleus, like Lead or Gold. This simplified view, however,
has to be confronted both with theoretical uncertainties and the actual
experimental feasibility.  Concerning the uncertainties, a simple
analysis shows that the dominant uncertainty coming from the scalar
matrix elements almost entirely cancels when taking ratios of conversion
rates (even using the conservative range $y \in [0,0.4]$ for the strange
scalar density matrix element).  Moreover, in the large-$Z$ tail of the
plot, some residual uncertainty arises from the input on the neutron
density profile.
When polarized proton scattering data exists, the uncertainty on the
ratios of conversion rates becomes negligible.  This point is
illustrated by Table~\ref{tab:targetdep}, where we report the detailed
breakdown of uncertainties in the ratios $B_{\mu \to e} ({\rm Ti})/B_{\mu \to
e} ({\rm Al})$ and $B_{\mu \to e} ({\rm Pb})/B_{\mu \to e} ({\rm Al})$.  For other
targets, the uncertainty induced by neutron densities never exceeds $5
\%$~\cite{ryuichiro}.
The conclusions of this exercise are that:
\begin{itemize}
\item The theoretical uncertainties 
(scalar matrix elements and neutron densities) largely cancel when we
      take a ratio.
\item As evident from Fig.~\ref{fig:target-dependence-1}, 
a realistic discrimination among models requires a measure of
$B_{\mu \to e} ({\rm Ti})/B_{\mu \to e} ({\rm Al})$ at the level of $5 \%$ or better, 
or alternatively  a measure of  $B_{\mu \to e} ({\rm Pb})/B_{\mu \to e} ({\rm Al})$ at the $20 \%$ level. 
These are two cases that well represent the trend in light and heavy
      target nuclei.
\end{itemize}

\begin{table}[t!]
\begin{center}
\begin{tabular}{|c|c|c|c|c|}
\hline 
  &  $S$   &   $D$  &  $V^{(\gamma)}$  & $V^{(Z)}$        
\\[5pt]
\hline 
%
$\frac{B(\mu \to e, {\rm Ti})}{B(\mu \to e, {\rm Al})}$     &  1.70 $\pm$ 0.005$_{y}$   & 
1.55 &  1.65  & 2.0  
\\[5pt]
\hline
$\frac{B(\mu \to e, {\rm Pb})}{B(\mu \to e, {\rm Al})}$     &  0.69 $\pm$ 0.02$_{\rho_n}$   & 
1.04  &  1.41  & 2.67  $\pm$ 0.06$_{\rho_n}$  
\\[5pt]
\hline
\end{tabular}
\end{center}
\caption{Ratios of  conversion rates in Titanium and Lead  over  Aluminum, 
in each of the four single-operator models:  scalar (S), dipole (D), vector 1 (photon coupling to the 
quarks)  and  vector 2 ($Z$ boson coupling to the quarks). 
In the scalar model, the scalar form factor induces a negligible uncertainty in the ratios 
involving two targets (denoted by the subscript $y$).   
In the case of Lead over Aluminum,  the small  uncertainty is dominated by the neutron density input 
(denoted by the subscript $\rho_n$).}
\label{tab:targetdep}
\end{table}

\section{Testing the  two-operator dominance hypothesis}
\label{sect:test2}

In the last section we have discussed how to test the hypothesis of a  single operator dominance, 
and how to discriminate among different single-operator dominance models. 
If the single operator dominance hypothesis fails,  one is lead to 
consider next simplest case, namely the two-operator dominance models,  
defined by the assumption that only two underlying 
operators have appreciable coefficients.  Each model is characterized by two parameters, 
the effective strength $C_{1}/\Lambda^2$ of one of the two operators 
and the ratio $C_{2}/C_{1}$ of the effective couplings of  the two dominant operators. 
This class of models can be tested so long as two double ratios of LFV rates 
are available (three LFV measurements!). 

For the sake of illustration, we will consider the following three two-operator models:
Dipole-Scalar, Dipole-Vector(Z) and Scalar-Vector(Z).   We consider both the 
case of constructive and destructive interference among the two dominant operators, 
assuming that the ratio of Wilson coefficients   $r \equiv C_{2}/C_{1} $ 
is real (a relative phase can be included but 
it would unnecessarily complicate the analysis at this early stage).
In order to test this class of models, one has to assume that at least three LFV 
processes have been observed, so one can construct two independent 
double ratios that are entirely determined by the single parameter $r$. 
In models involving  the Dipole operator among the dominant terms 
(such as Dipole-Scalar and Dipole-Vector)   
the three observables could be  (i)  $\mu \rightarrow e \gamma$ and  $\mu \to e$ conversion 
in two different targets;  or  (ii)  $\mu \to e$ conversion in three different targets. 
In models that do not involve a Dipole term (such as Scalar-Vector),  only the 
possibility (ii) above is available. 
As representative target nuclei, we have chosen aluminum (${\rm Al}$), 
titanium (${\rm Ti}$), and lead (${\rm Pb}$).

\subsection{Dipole-Scalar}

In terms of the parameters introduced  in Section~\ref{sect:models1},  this model is defined by 
$C_{S} \neq 0$ and  $C_D \equiv \pm \frac{r}{8e} \; C_S$.
The single-operator models are recovered in the limiting cases $r \to 0$ (scalar) and  
$r \to \infty$  (dipole)~\footnote{We  consider here  the case in which  dipole and scalar operators 
produce  outgoing lepton with definite chirality (L or R). 
If both chiralities are allowed,  then in principle   $C_{DR}/C_{SR}  \neq C_{DL}/C_{SL}$ 
and one more parameter has to be introduced in the analysis.}. 
Note that in this particular case the asymptotic dipole regime is reached already for 
$r  \ll  1$ because of the peculiar normalization of the scalar operators (suppressed 
by the factor $G_F m_q m_\mu$). 
\begin{figure}[H]
\begin{center}
\begin{minipage}{.49\linewidth}
\centering 
\includegraphics[width=7.5cm]{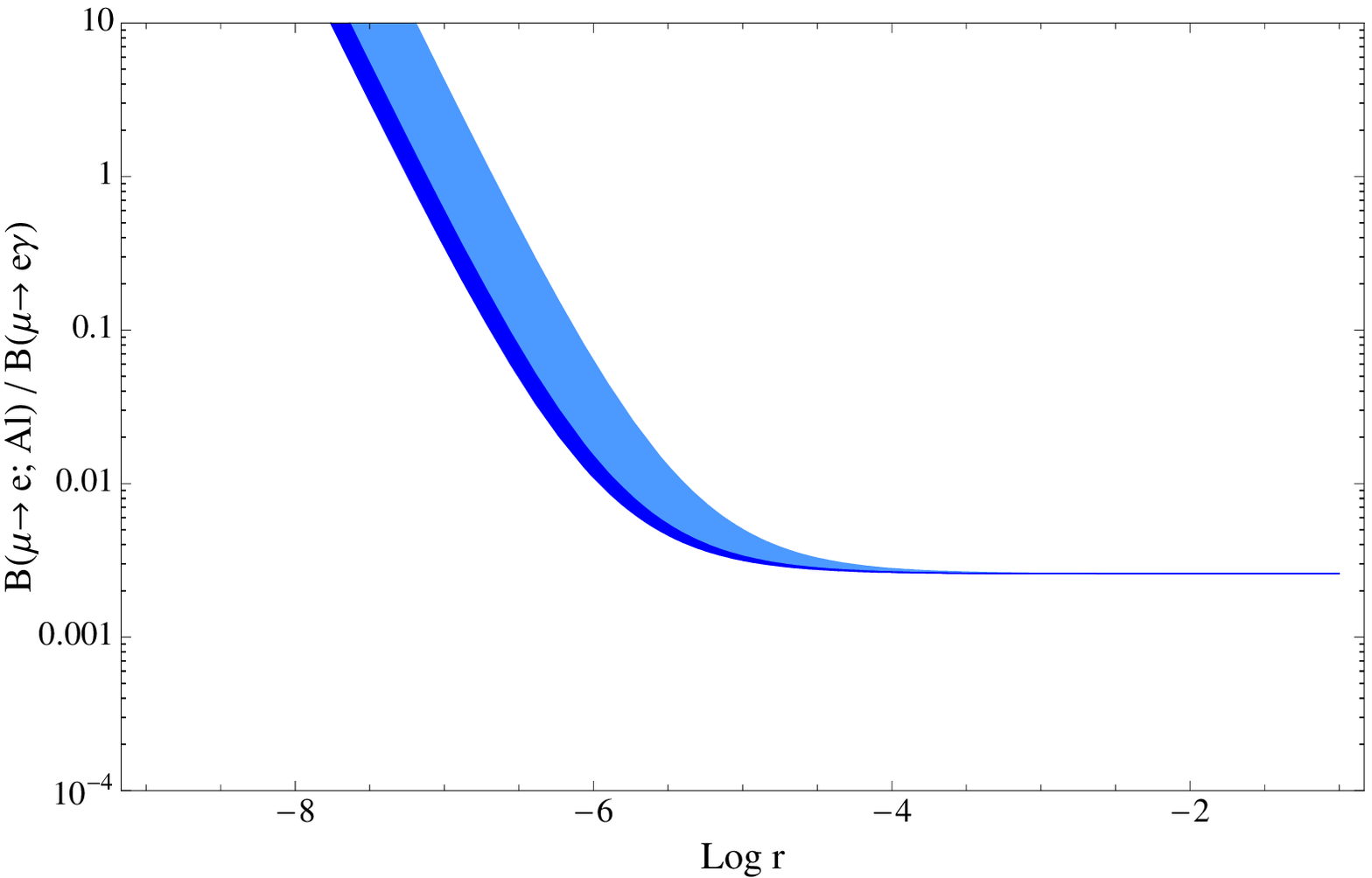}
\center{(a)}
\end{minipage}
\end{center}
\begin{minipage}{.49\linewidth}
\centering 
\includegraphics[width=7.5cm]{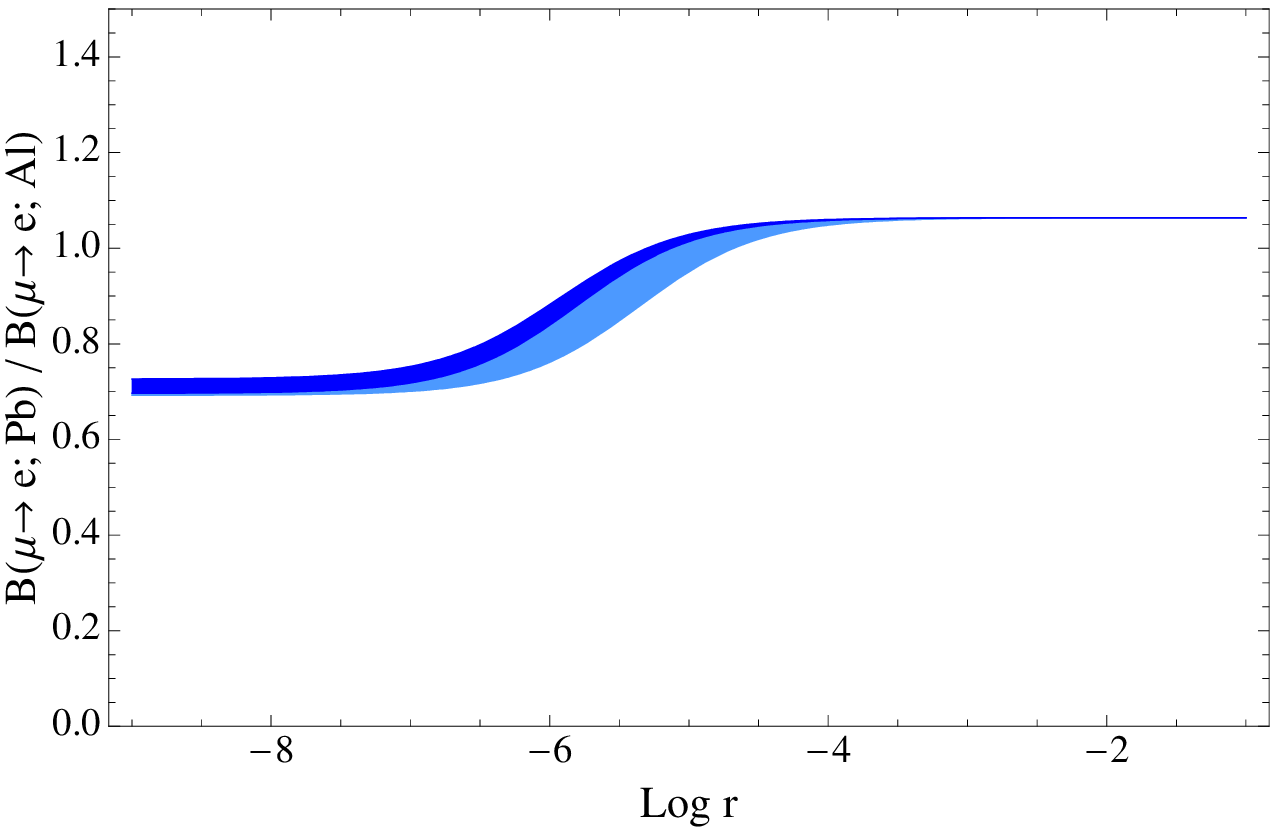}
\center{(b)}
\end{minipage}
\begin{minipage}{.49\linewidth}
\centering
\includegraphics[width=7.5cm]{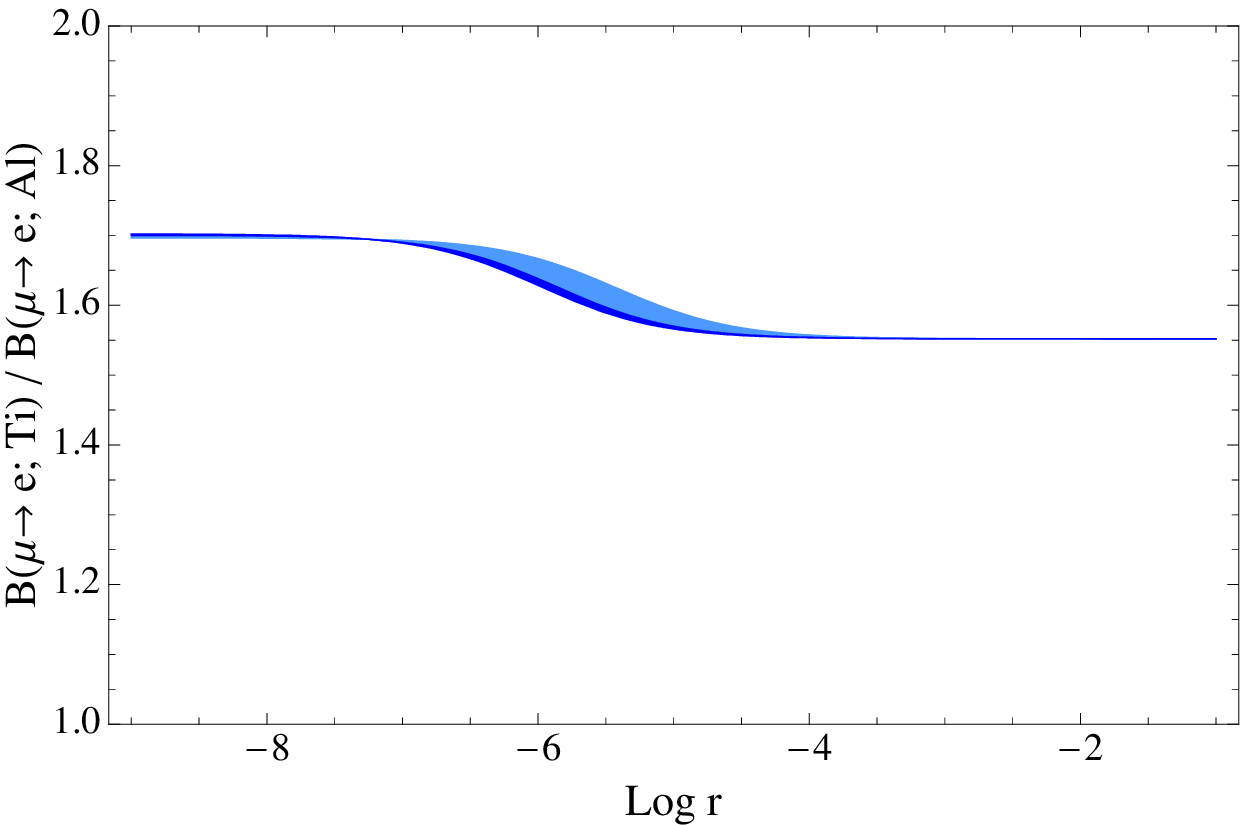}
\center{(c)}
\end{minipage}
\caption{{\it Dipole-Scalar model}: 
Ratios  $B_{\mu \to e} ({\rm Al})/B_{\mu \to e \gamma}$ (panel (a)), 
$B_{\mu \to e} ({\rm Pb})/B_{\mu \to e} ({\rm Al})$ (panel (b)), and 
$B_{\mu \to e} ({\rm Ti})/B_{\mu \to e} ({\rm Al})$ (panel (c))  
as a function of ${\rm Log}_{10} (r)$  
for positive $C_D/C_S$.
See text for details.}
\label{figuno}
\end{figure}

\begin{figure}[H]
\begin{center}
\begin{minipage}{.49\linewidth}
\centering 
\includegraphics[width=7.5cm]{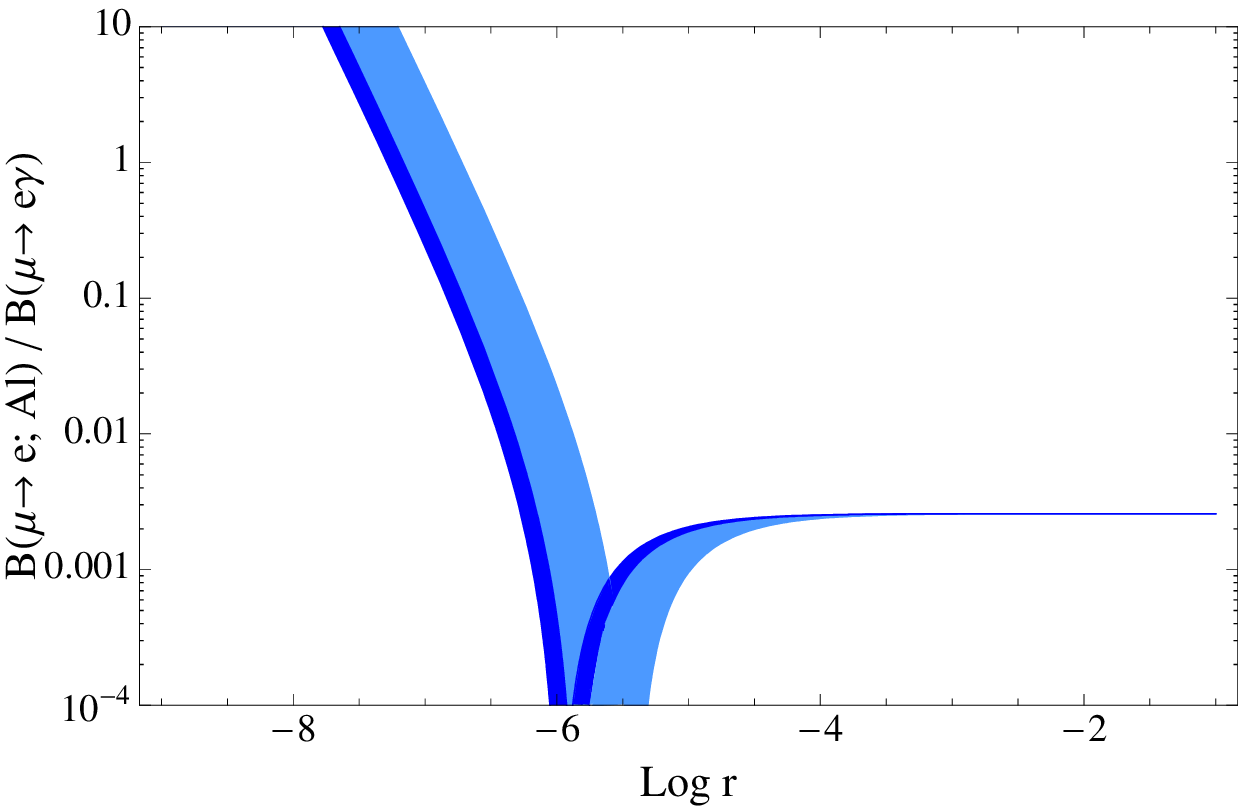}
\center{(a)}
\end{minipage}
\end{center}
\begin{minipage}{.49\linewidth}
\centering 
\includegraphics[width=7.5cm]{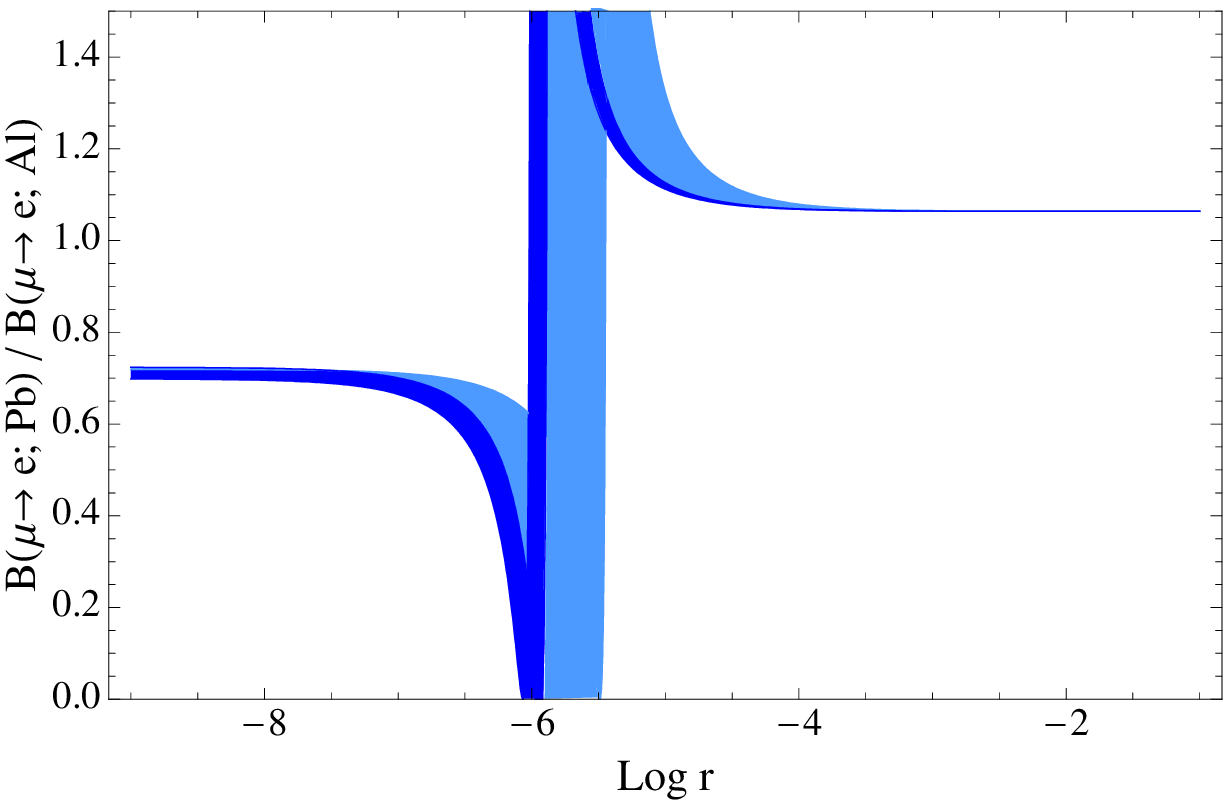}
\center{(b)}
\end{minipage}
\begin{minipage}{.49\linewidth}
\centering 
\includegraphics[width=7.5cm]{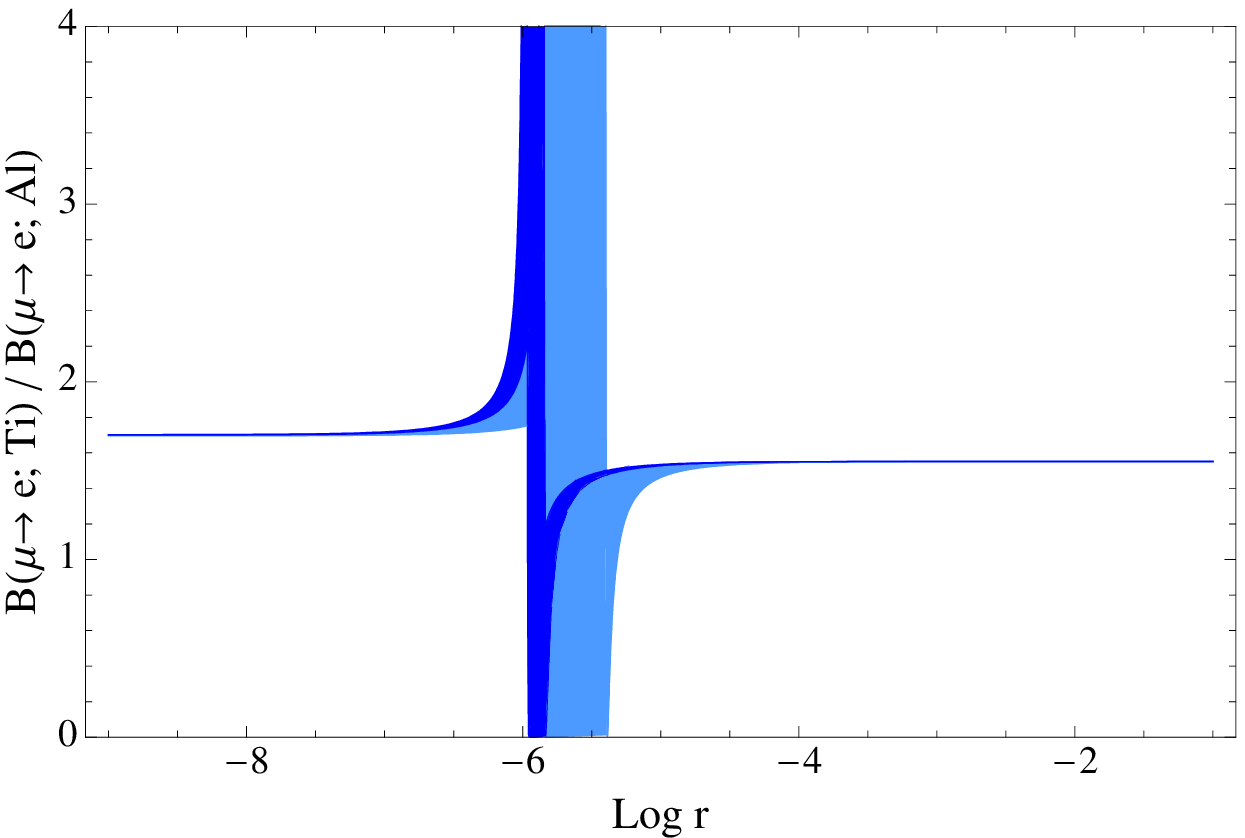}
\center{(c)}
\end{minipage}
\caption{{\it Dipole-Scalar model}: 
Ratios  $B_{\mu \to e} ({\rm Al})/B_{\mu \to e \gamma}$  (panel (a)), 
$B_{\mu \to e} ({\rm Pb})/B_{\mu \to e} ({\rm Al})$ (panel (b)), and 
$B_{\mu \to e} ({\rm Ti})/B_{\mu \to e} ({\rm Al})$ (panel (c)) 
as a function of ${\rm Log}_{10} (r)$  
for negative $C_D/C_S$.
See text for details.}
\label{figunoreves}
\end{figure}

We illustrate the features of this model in Figs. \ref{figuno} and
\ref{figunoreves}, which correspond to positive and negative sign of the
ratio $C_D/C_S$, respectively.  Panel (a) shows the behavior of $B_{\mu
\to e} ({\rm Al})/B_{\mu \to e \gamma}$ versus the parameter $r$, while panels
(b) and (c) show the ratios $B_{\mu \to e} ({\rm Pb})/B_{\mu \to e} ({\rm Al})$ and
$B_{\mu \to e} ({\rm Ti})/B_{\mu \to e} ({\rm Al})$, respectively.
In panels (a) and (c)  the curve is widened in the interference region by the uncertainty in the 
scalar form  factors.  The dominant uncertainty comes from the input parameter $y$, characterizing 
the strangeness content of the nucleon. 
On the other hand, the ratio $B_{\mu \to e} ({\rm Pb})/B_{\mu \to e} ({\rm Al})$ is
affected not only by the uncertainty in the scalar form factors, but
also by the uncertainty induced in the overlap integral by the neutron
density in Pb.  The width of the bands in panel (b) is determined by the
most conservative combination of two kind of uncertainties.

In all panels the wide band corresponds to the range $y\in [0,0.4]$,
while the narrow band corresponds to the range $y \in [0,0.05]$.  This
illustrates the effect of current and future hadronic uncertainties on
the process of extracting information on short distance LFV couplings.
The prominent feature in Fig.~\ref{figunoreves} is induced by the destructive interference 
dipole and scalar amplitudes. 

\subsection{Dipole-Vector}

\begin{figure}[H]
\begin{center}
\begin{minipage}{.49\linewidth}
\centering 
\includegraphics[width=7.5cm]{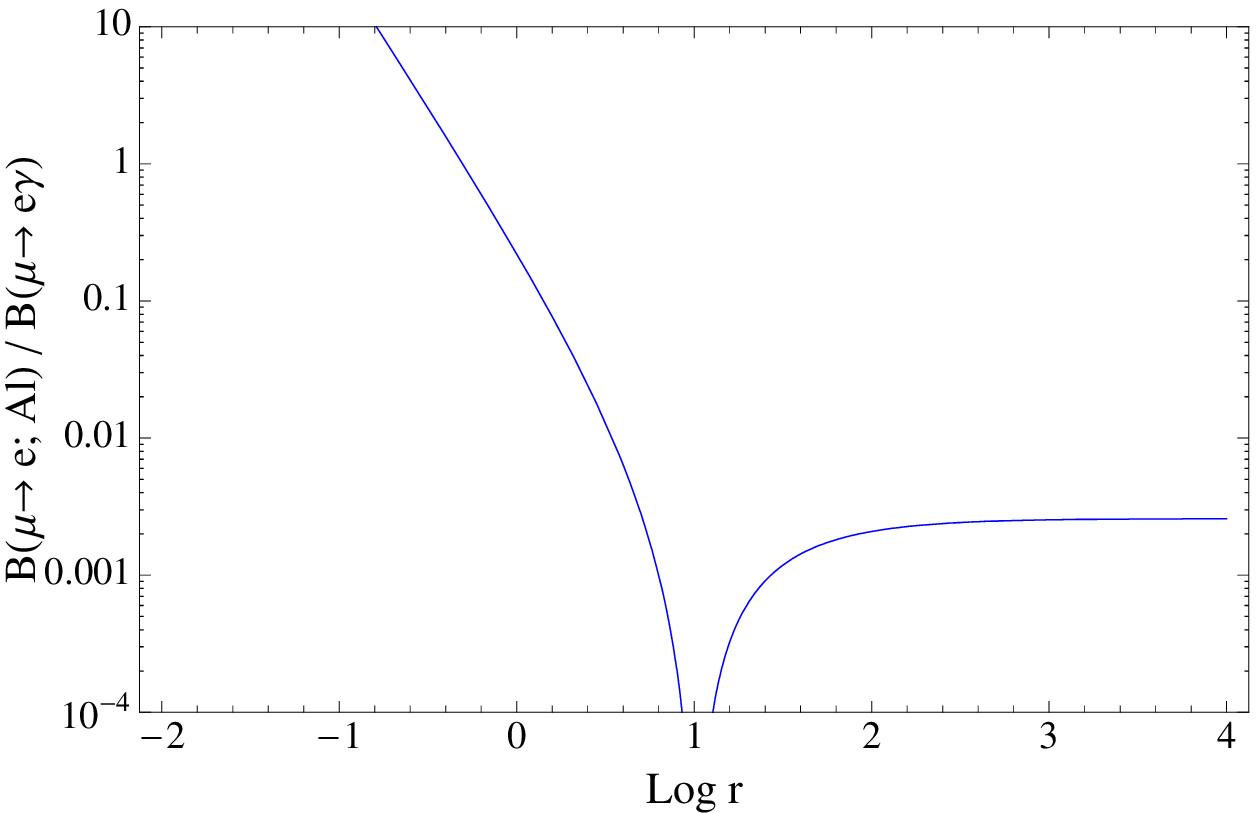}
\center{(a)}
\end{minipage}
\end{center}
\begin{minipage}{.49\linewidth}
\centering 
\includegraphics[width=7.5cm]{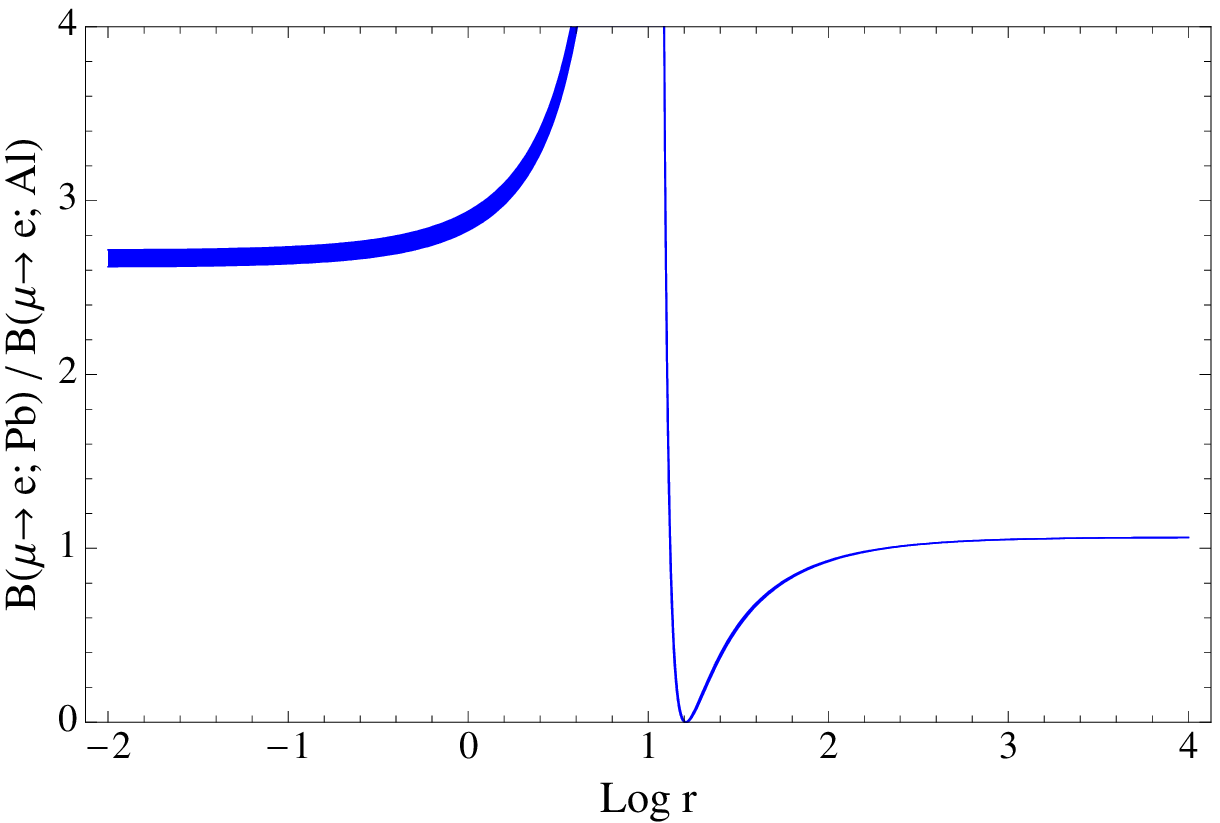}
\center{(b)}
\end{minipage}
\begin{minipage}{.49\linewidth}
\centering 
\includegraphics[width=7.5cm]{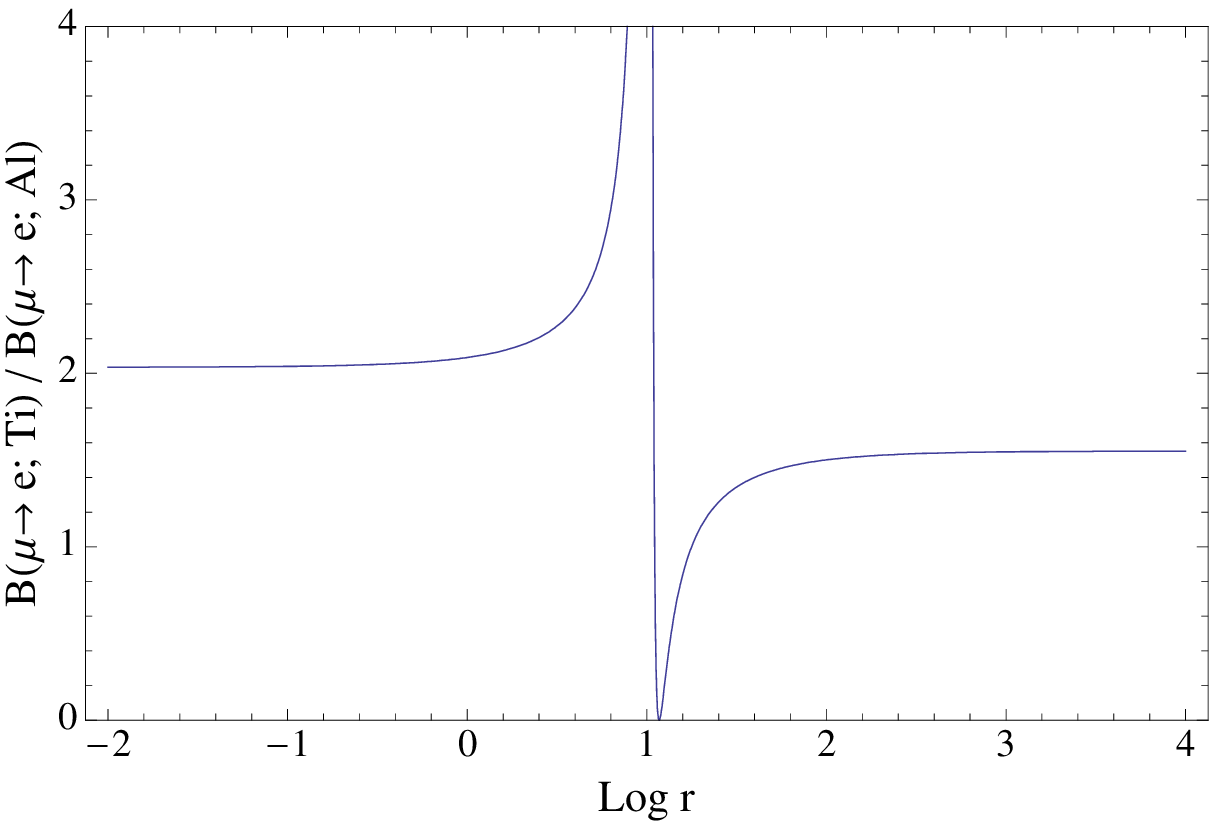}
\center{(c)}
\end{minipage}
\caption{ {\it Dipole-Vector model}: 
Ratios  $B_{\mu \to e} ({\rm Al})/B_{\mu \to e \gamma}$ (panel (a)), 
$B_{\mu \to e} ({\rm Pb})/B_{\mu \to e} ({\rm Al})$ (panel (b)), and 
$B_{\mu \to e} ({\rm Ti})/B_{\mu \to e} ({\rm Al})$ (panel (c))  
as a function of ${\rm Log}_{10} (r)$  
for positive $C_D/C_V$. 
See text for details.}
\label{figdos}
\end{figure}

\begin{figure}[H]
\begin{center}
\begin{minipage}{.49\linewidth}
\centering 
\includegraphics[width=7.5cm]{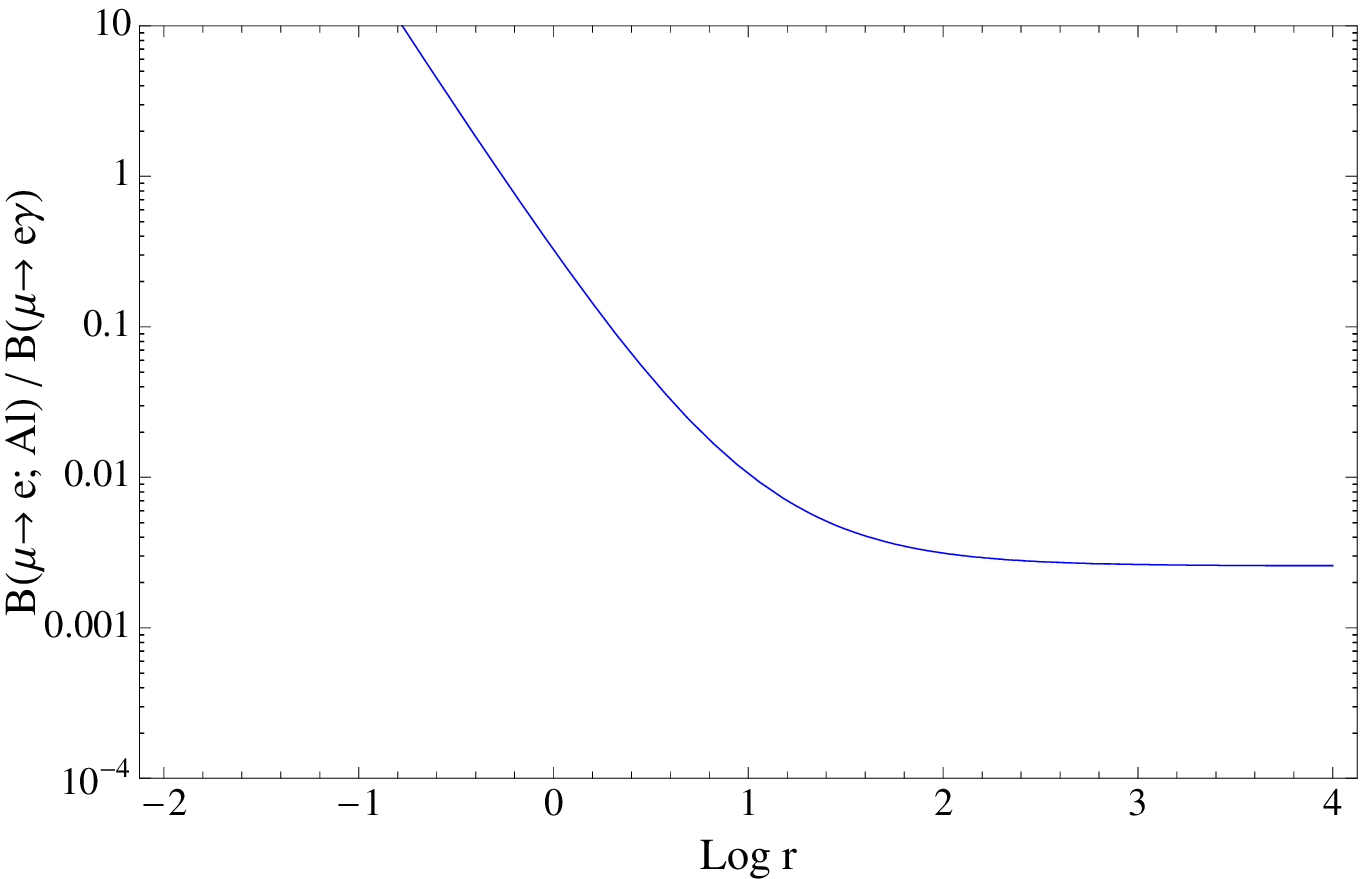}
\center{(a)}
\end{minipage}
\end{center}
\begin{minipage}{.49\linewidth}
\centering 
\includegraphics[width=7.5cm]{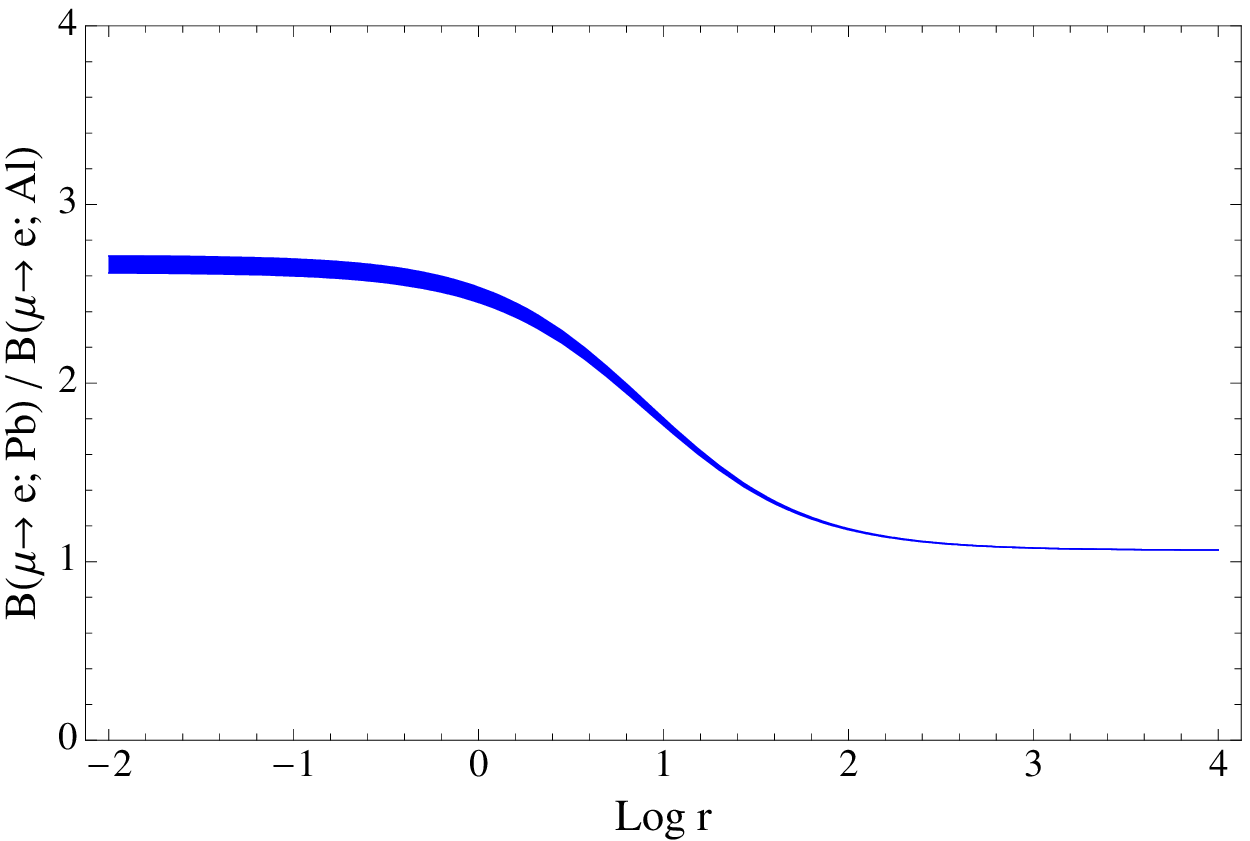}
\center{(b)}
\end{minipage}
\begin{minipage}{.49\linewidth}
\centering 
\includegraphics[width=7.5cm]{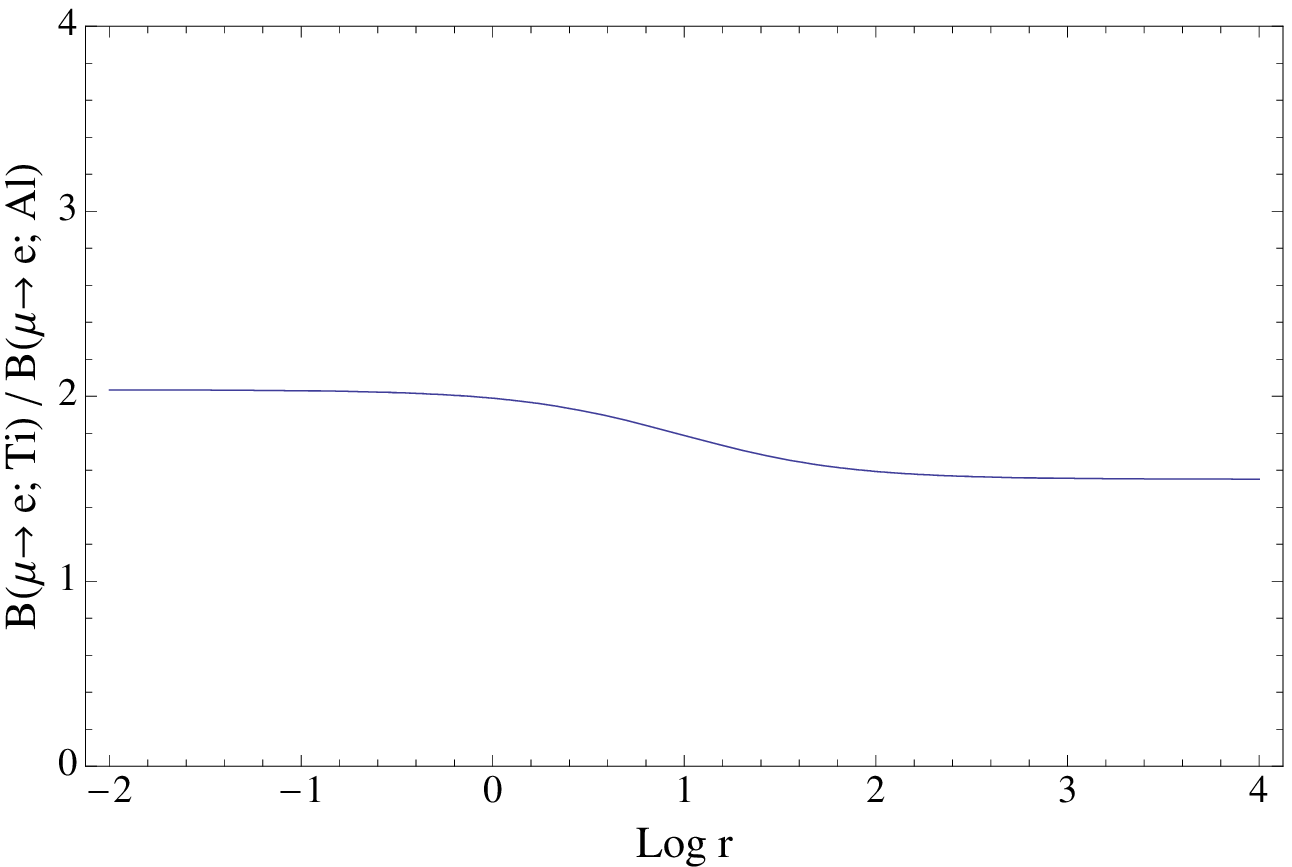}
\center{(c)}
\end{minipage}
\caption{ {\it Dipole-Vector model}: 
Ratios  $B_{\mu \to e} ({\rm Al})/B_{\mu \to e \gamma}$ (panel (a)), 
$B_{\mu \to e} ({\rm Pb})/B_{\mu \to e} ({\rm Al})$ (panel (b)), and 
$B_{\mu \to e} ({\rm Ti})/B_{\mu \to e} ({\rm Al})$ (panel (c))  
as a function of ${\rm Log}_{10} (r)$  
for negative $C_D/C_V$.  
See text for details.}
\label{figdosreves}
\end{figure}

In terms of the parameters defined in Section~\ref{sect:models1},  this model is defined by 
$C_V \neq 0$ and  $C_D \equiv \pm \frac{r}{8e}\; C_V$. 
The single-operator models are recovered in the limiting cases $r \to 0$ (vector) and  
$r \to \infty$  (dipole). 
In figures  \ref{figdos}  and  \ref{figdosreves}  we plot  
the ratios   $B_{\mu \to e} ({\rm Al})/B_{\mu \to e \gamma}$ (panel (a)), 
$B_{\mu \to e} ({\rm Pb})/B_{\mu \to e} ({\rm Al})$  (panel (b)), and 
$B_{\mu \to e} ({\rm Ti})/B_{\mu \to e} ({\rm Al})$ (panel (c))
versus the  parameter $r$. 
Figures  \ref{figdos}  and  \ref{figdosreves}  correspond to 
positive and negative sign of the ratio $C_D/C_V$, respectively. 
Within this model, the only source of uncertainty arises from the 
vector overlap integral $V^{(n)}({\rm Pb})$, sensitive to the neutron density in Pb. 
This uncertainty is quantified by the thickness of the band in panel (b).

\subsection{Scalar-Vector}

\begin{figure}[H]
\begin{minipage}{.49\linewidth}
\centering 
\includegraphics[width=7.5cm]{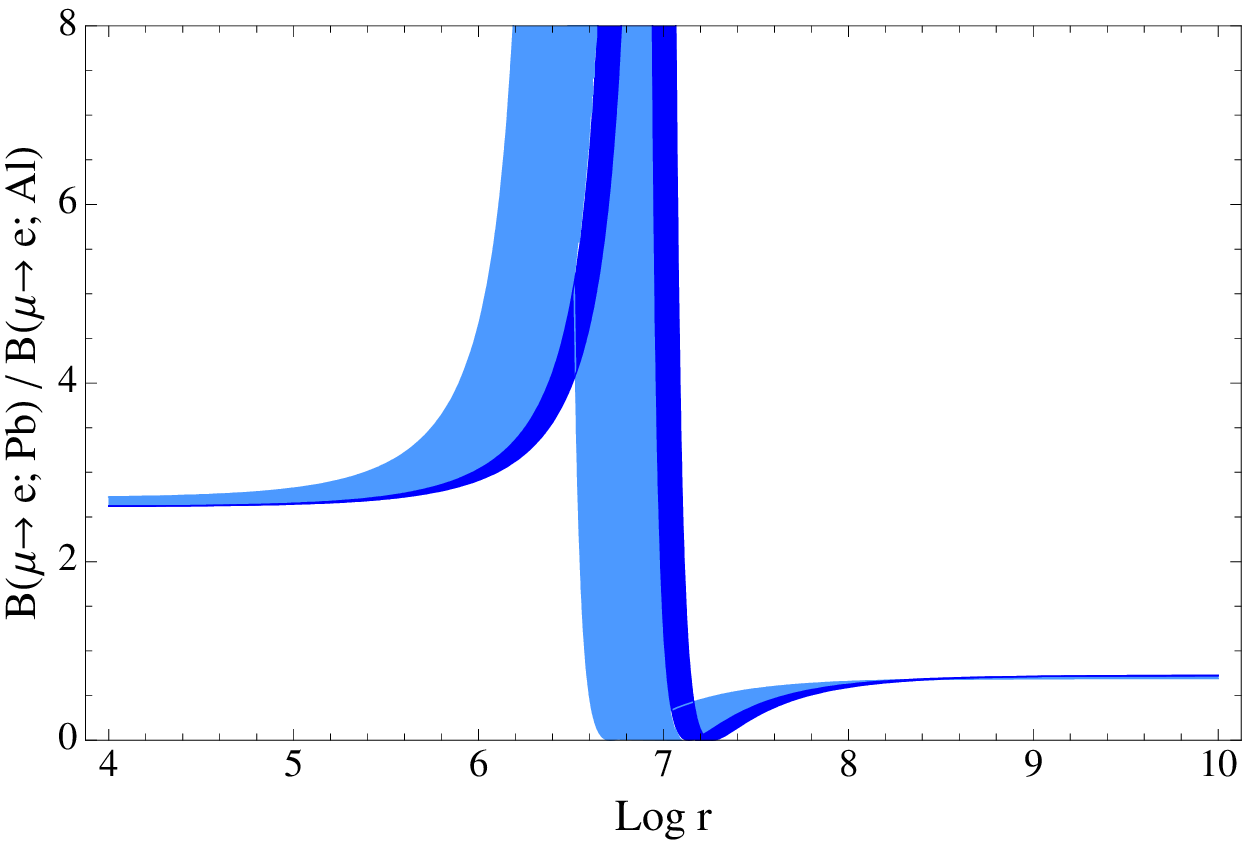}
\center{(b)}
\end{minipage}
\begin{minipage}{.49\linewidth}
\centering 
\includegraphics[width=7.5cm]{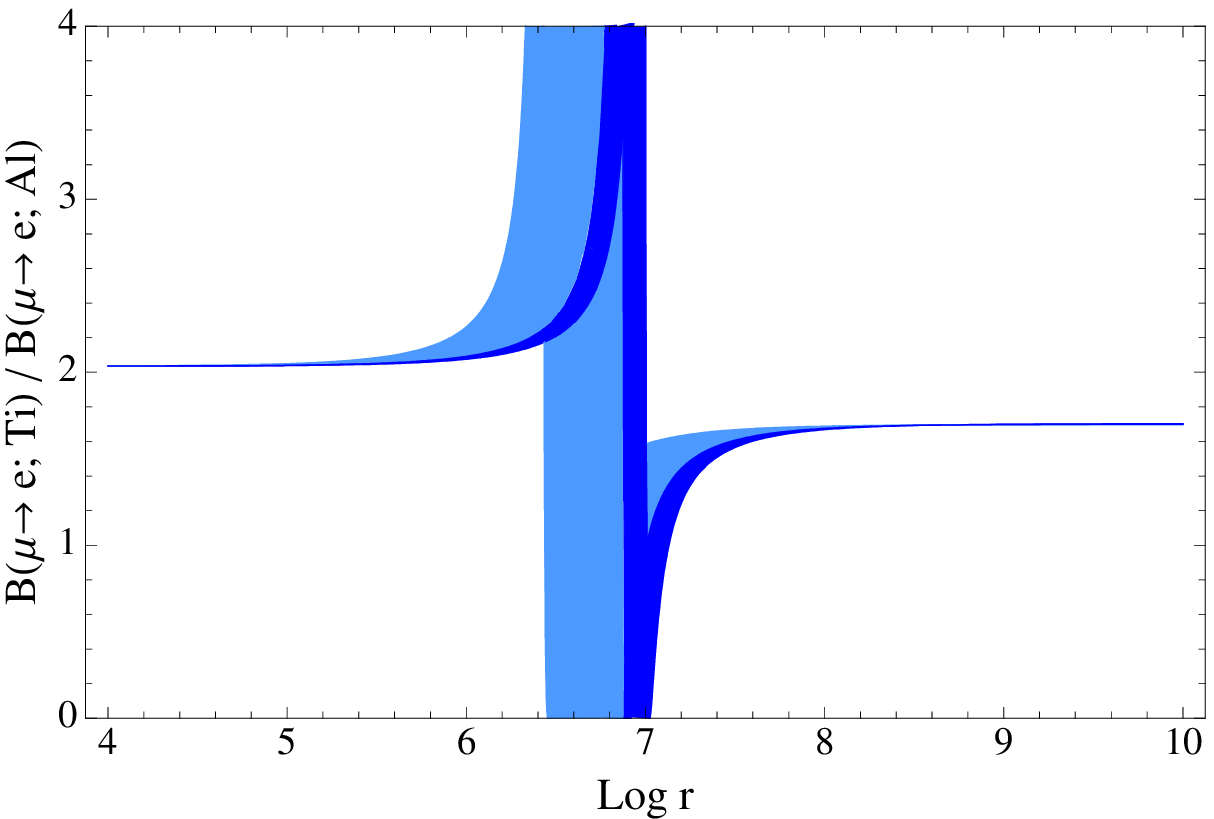}
\center{(c)}
\end{minipage}
\caption{
{\it Scalar-Vector model}: 
Ratios   $B_{\mu \to e} ({\rm Pb})/B_{\mu \to e} ({\rm Al})$ (panel (b)) and 
$B_{\mu \to e} ({\rm Ti})/B_{\mu \to e} ({\rm Al})$ (panel (c)) 
as a function of ${\rm Log}_{10} (r)$  
for positive $C_S/C_V$.
See text for details.
}
\label{figtres}
\end{figure}

\begin{figure}[H]
\begin{minipage}{.49\linewidth}
\centering 
\includegraphics[width=7.5cm]{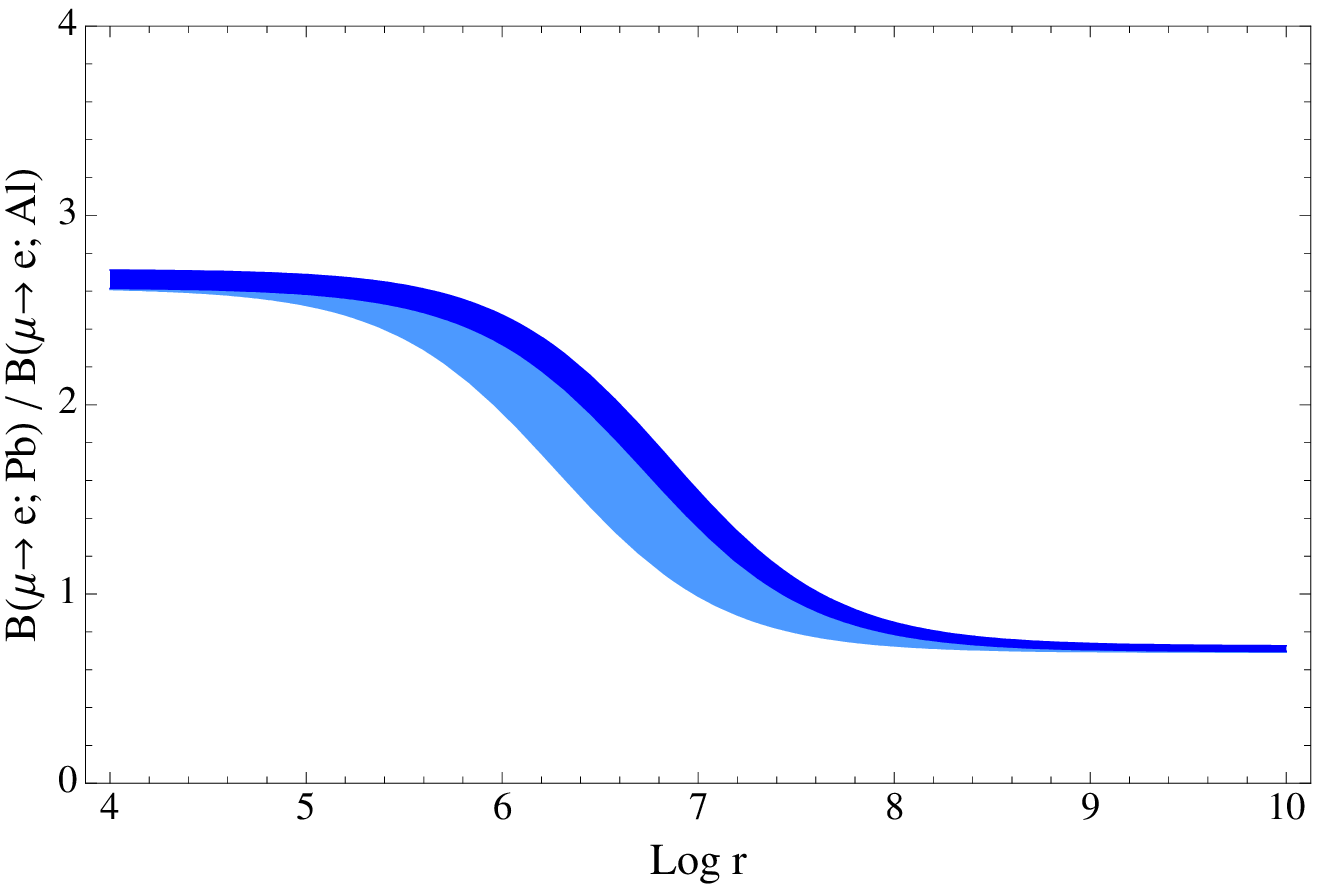}
\center{(b)}
\end{minipage}
\begin{minipage}{.49\linewidth}
\centering 
\includegraphics[width=7.5cm]{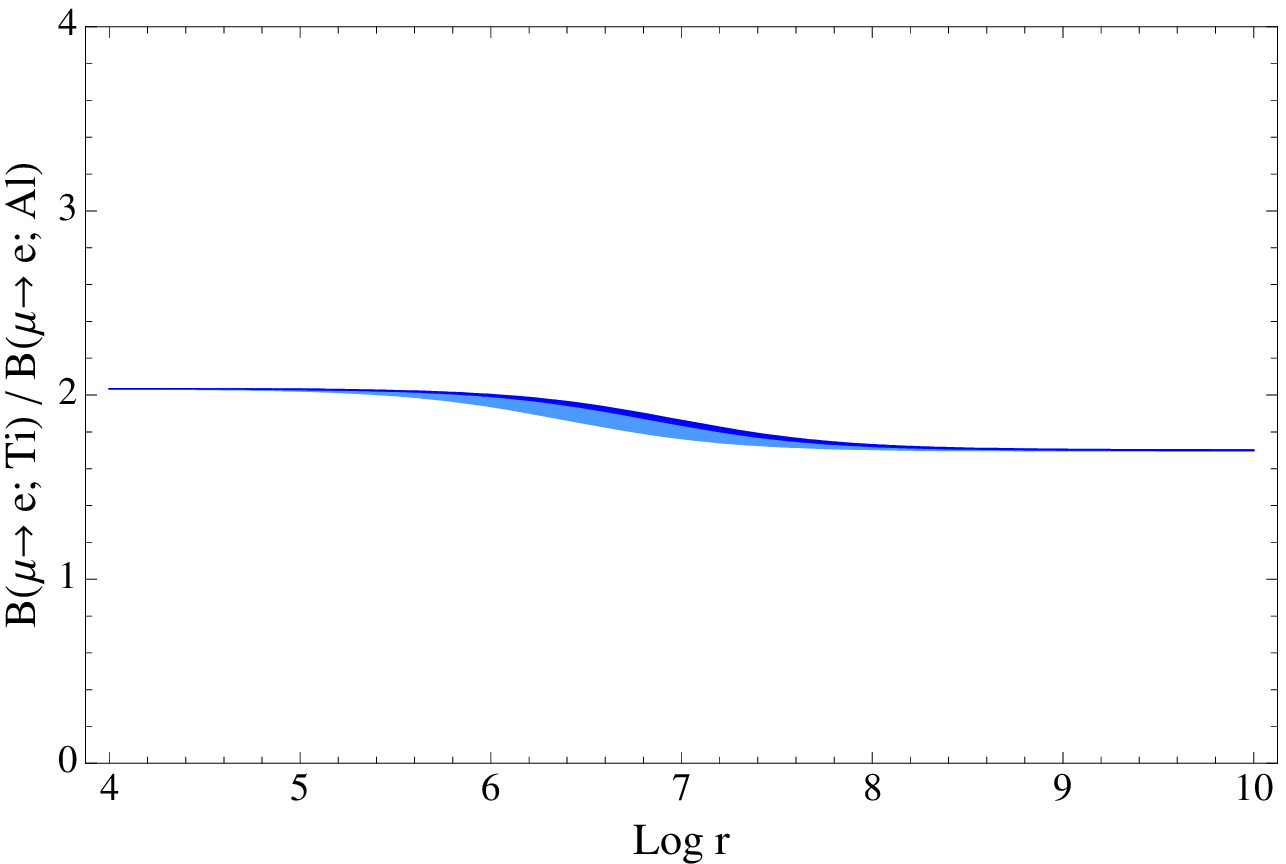}
\center{(c)}
\end{minipage}
\caption{
{\it Scalar-Vector model}: 
Ratios  $B_{\mu \to e} ({\rm Pb})/B_{\mu \to e} ({\rm Al})$ (panel (b)) and 
$B_{\mu \to e} ({\rm Ti})/B_{\mu \to e} ({\rm Al})$ (panel (c)) 
as a function of ${\rm Log}_{10} (r)$  
for negative $C_S/C_V$.
See text for details.}
\label{figtresreves}
\end{figure}

In terms of the parameters defined in Section~\ref{sect:models1},  this model is defined by 
$C_V \neq 0$ and  $C_S \equiv \pm r \,  C_V$. 
The single-operator models are recovered in the limiting cases $r \to 0$ (vector) and  
$r \to \infty$  (scalar). 
Since the Dipole term is assumed to be subdominant, in this case we include in the analysis 
only the ratios  $B_{\mu \to e} ({\rm Pb})/B_{\mu \to e} ({\rm Al})$  and  $B_{\mu \to e} ({\rm Ti})/B_{\mu \to e} ({\rm Al})$, 
shown in panels (b) and (c)  of Figures~\ref{figtres} and \ref{figtresreves} 
(for positive and negative values of $C_S/C_V$, respectively).  
While the ratio  $B_{\mu \to e} ({\rm Ti})/B_{\mu \to e} ({\rm Al})$  is affected only by 
the uncertainty in $y$,  the ratio $B_{\mu \to e} ({\rm Pb})/B_{\mu \to e} ({\rm Al})$  is affected 
also  by the uncertainty in the Pb  neutron density (through the overlap integrals). 
The width of the bands in the plots is determined  
by the most conservative combination of two kind of uncertainties.  

In all panels   the wide band corresponds to the range $y\in [0,0.4]$, while  
the narrow band corresponds to the range $y \in [0,0.05]$. 
As in the case of the Dipole-Scalar model, 
the bands  illustrate the effect of current and future  hadronic uncertainties on
extracting  short distance LFV couplings.

We conclude this section by summarizing what one could learn about 
the two-operator dominance models in the case that two double ratios 
of LFV rates could be measured experimentally. 
Our exercise shows that: 
\begin{itemize}

\item The current theoretical uncertainty on the strange content of the nucleon
prevents a realistic test of  the two-operator models involving the Scalar amplitude. 
The range $y \in [0,0.4]$ induces uncertainties of up to one order of magnitude 
in the relevant double ratios in the interference region (thick  bands in all plots above).
However, the uncertainty within reach of lattice QCD calculations will remove this 
obstacle in the coming years (this is illustrated by the thin bands in all plots above).  

\item  Testing and discriminating among two-operator dominance models  
requires an experimental precision on the LFV rates that is comparable to the 
one needed to test the single operator models. 

\end{itemize}

\section{Application to a SUSY model}
\label{sect:susy}

An example of the two-operator dominance model is given by 
a SUSY scenario with flavor mixing in the left-handed sleptons.  
Such a mixing, for example, can be induced from the Yukawa
interaction in the see-saw model.
As it is shown in Ref.~\cite{largetanbeta1}, the scalar operator
originated from the Higgs-boson-exchange diagrams can be sizable in this
model if $\tan \beta$ is large and the heavy Higgs boson is relatively
lighter than the other SUSY particles. The ratio $B_{\mu \to e} ({\rm
Al}) / B_{\mu \to e \gamma}$ can therefore be enhanced in such a
parameter region, while the  ratio $B_{\mu \to e} ({\rm Pb}) / B_{\mu \to e} ({\rm Al})$ 
can show substantial deviations from the dipole-dominance value. 

\begin{figure}[t]
\begin{minipage}{.49\linewidth}
\centering 
\includegraphics[width=7cm]{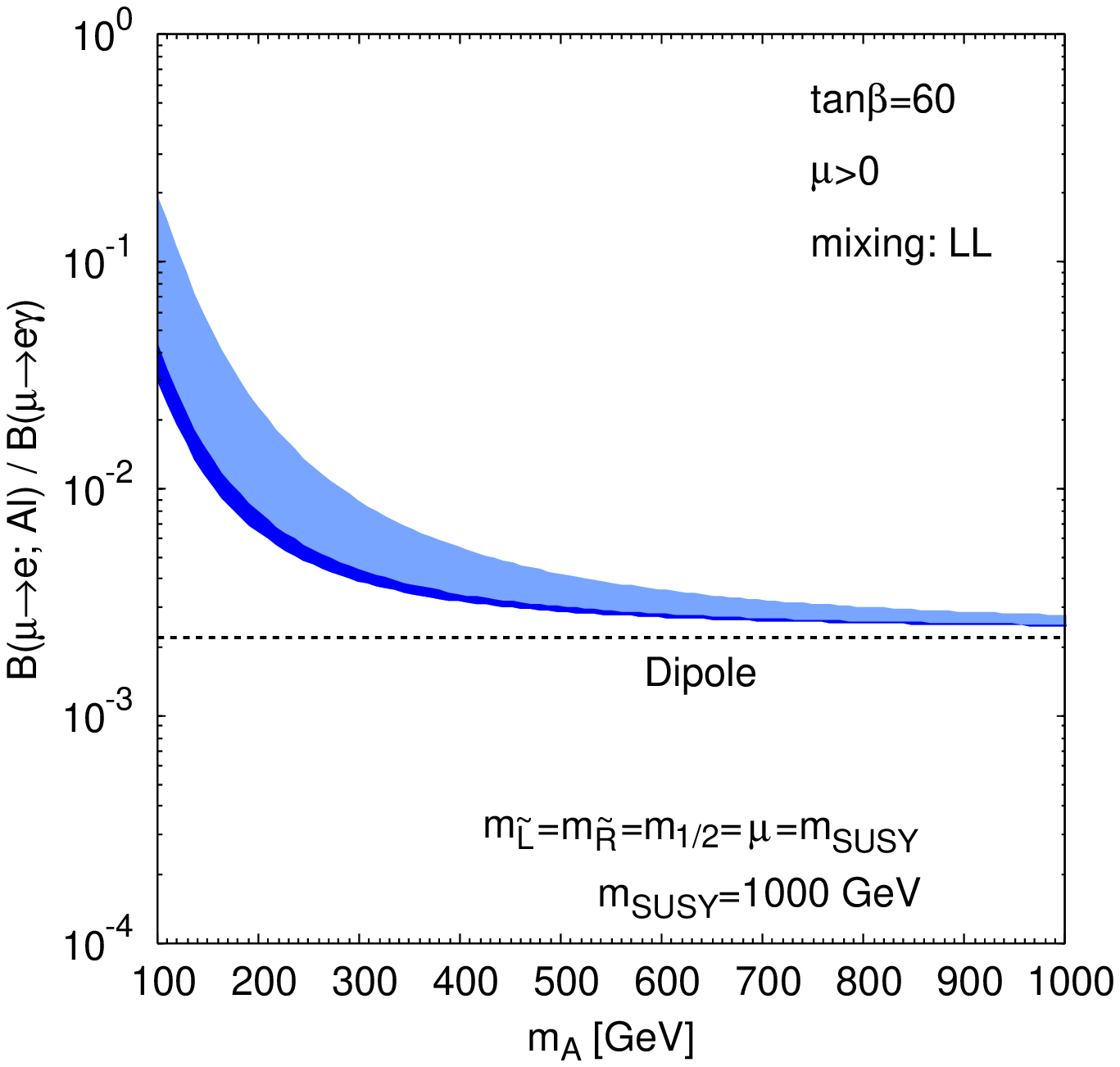}
\end{minipage}
\begin{minipage}{.49\linewidth}
\centering 
\includegraphics[width=7cm]{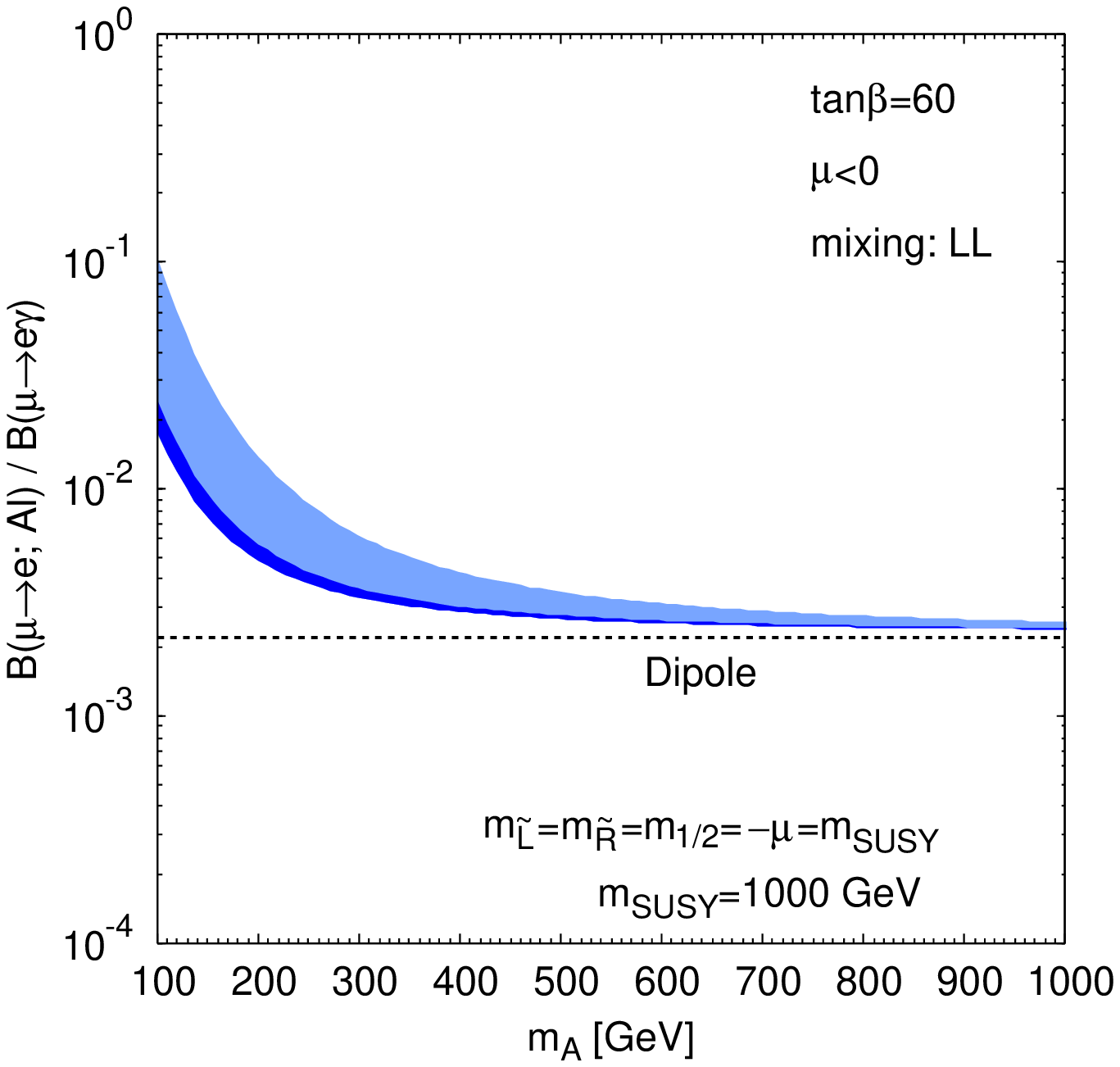}
\end{minipage}
\caption{ The pseudoscalar-Higgs mass dependence of $B_{\mu \to e} ({\rm
Al}) / B_{\mu \to e \gamma}$ in a SUSY model with a left-handed slepton
mixing. The slepton masses, the gaugino masses at the GUT scale, and the
Higgsino mass parameter are all fixed to be 1~TeV. The light and dark
shaded regions respectively correspond to the conservative and
optimistic ranges of the $y$ parameter.  } \label{fig:SUSY}
\end{figure}

\begin{figure}[t]
\begin{minipage}{.49\linewidth}
\centering 
\includegraphics[width=7cm]{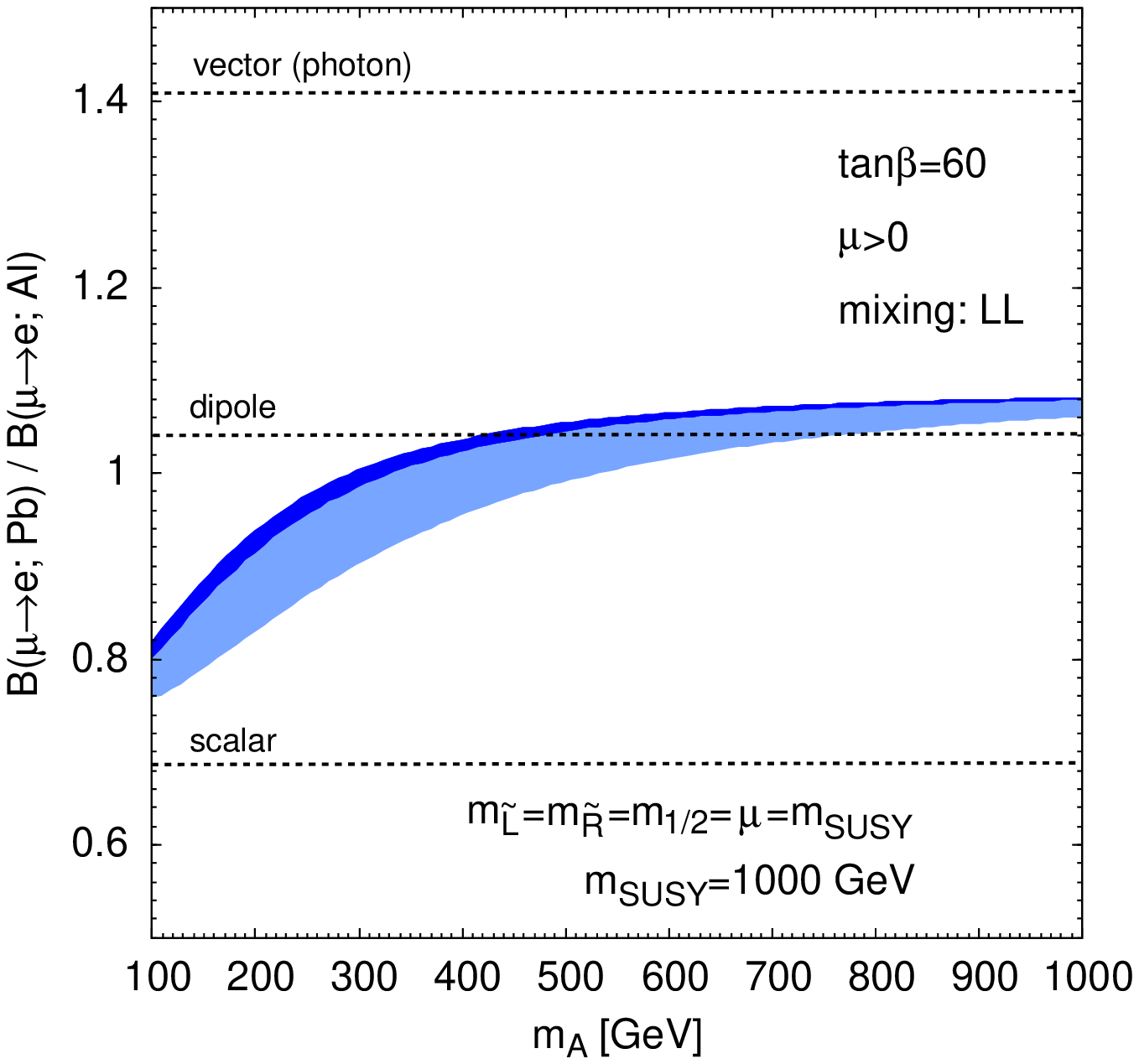}
\end{minipage}
\begin{minipage}{.49\linewidth}
\centering 
\includegraphics[width=7cm]{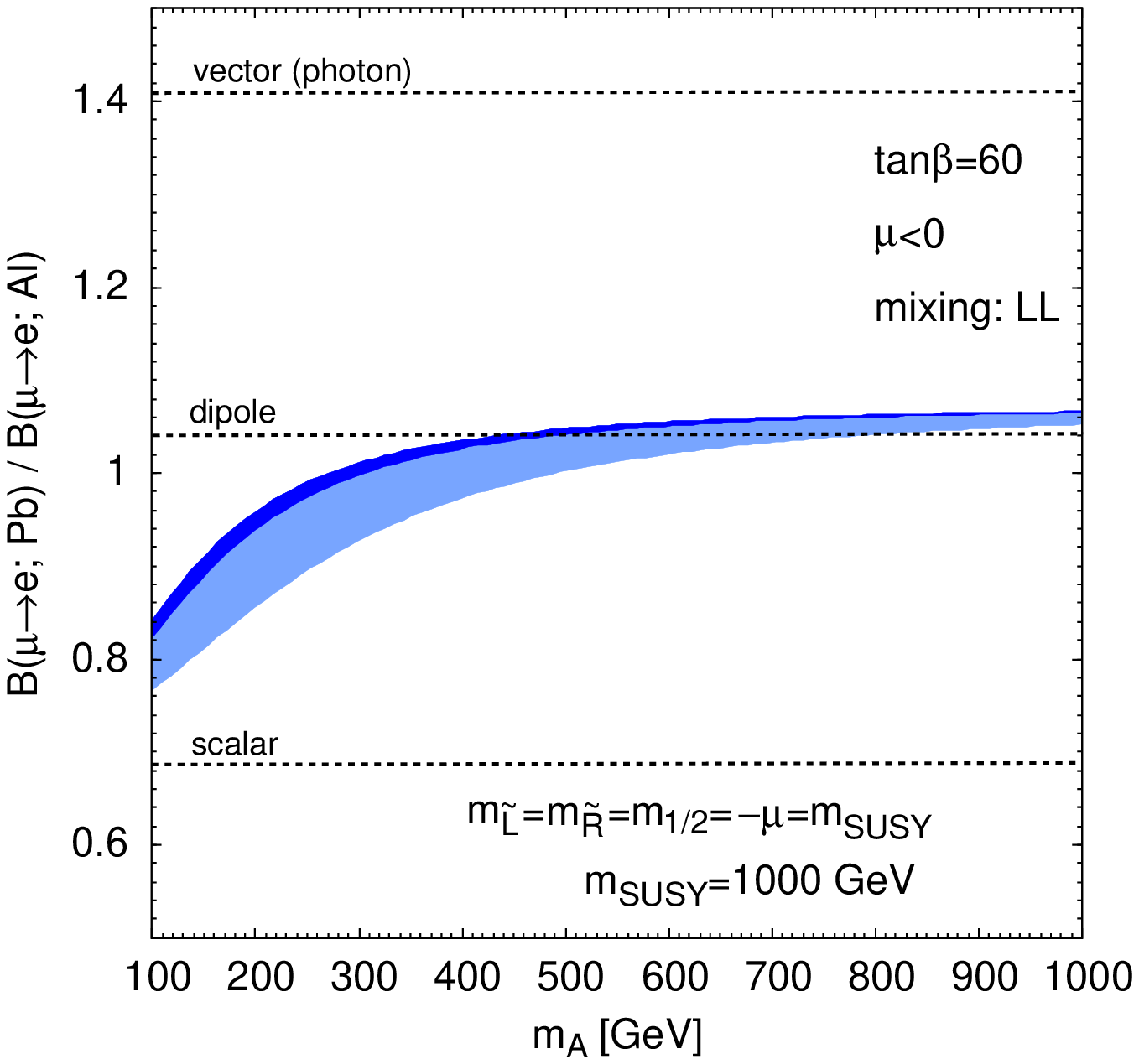}
\end{minipage}
\caption{ The pseudoscalar-Higgs mass dependence of $B_{\mu \to e} ({\rm
Pb}) / B_{\mu \to e} ({\rm Al})$ in a SUSY model with a left-handed slepton
mixing. The slepton masses, the gaugino masses at the GUT scale, and the
Higgsino mass parameter are all fixed to be 1~TeV. The light and dark
shaded regions respectively correspond to the conservative and
optimistic ranges of the $y$ parameter.  } \label{fig:SUSY2}
\end{figure}

In Fig.~\ref{fig:SUSY} we show the pseudoscalar-Higgs mass ($m_A$)
dependence of the ratio for $\mu > 0$ (left) and $\mu < 0$ (right).  We
have taken the common mass ($m_{\rm SUSY} = 1$~TeV) for the slepton
masses, the universal gaugino mass at the GUT scale, and the Higgsino mass
parameter, and we fixed $\tan \beta = 60$. Since the scalar operator
does not decouple in the $m_{\rm SUSY} \to \infty$ limit, we see the
enhancement in the small $m_A$ region.
The light (dark) shaded regions correspond to the conservative
(optimistic) range of the $y$ parameter, $y \in [0,0.4]$ ($y \in
[0,0.05]$).
Within the same framework,  the ratio $B_{\mu \to e} ({\rm Pb}) / B_{\mu \to e} ({\rm Al})$ 
is shown in Fig.~\ref{fig:SUSY2}. 

\begin{figure}[t]
\begin{minipage}{.49\linewidth}
\centering 
\includegraphics[width=7cm]{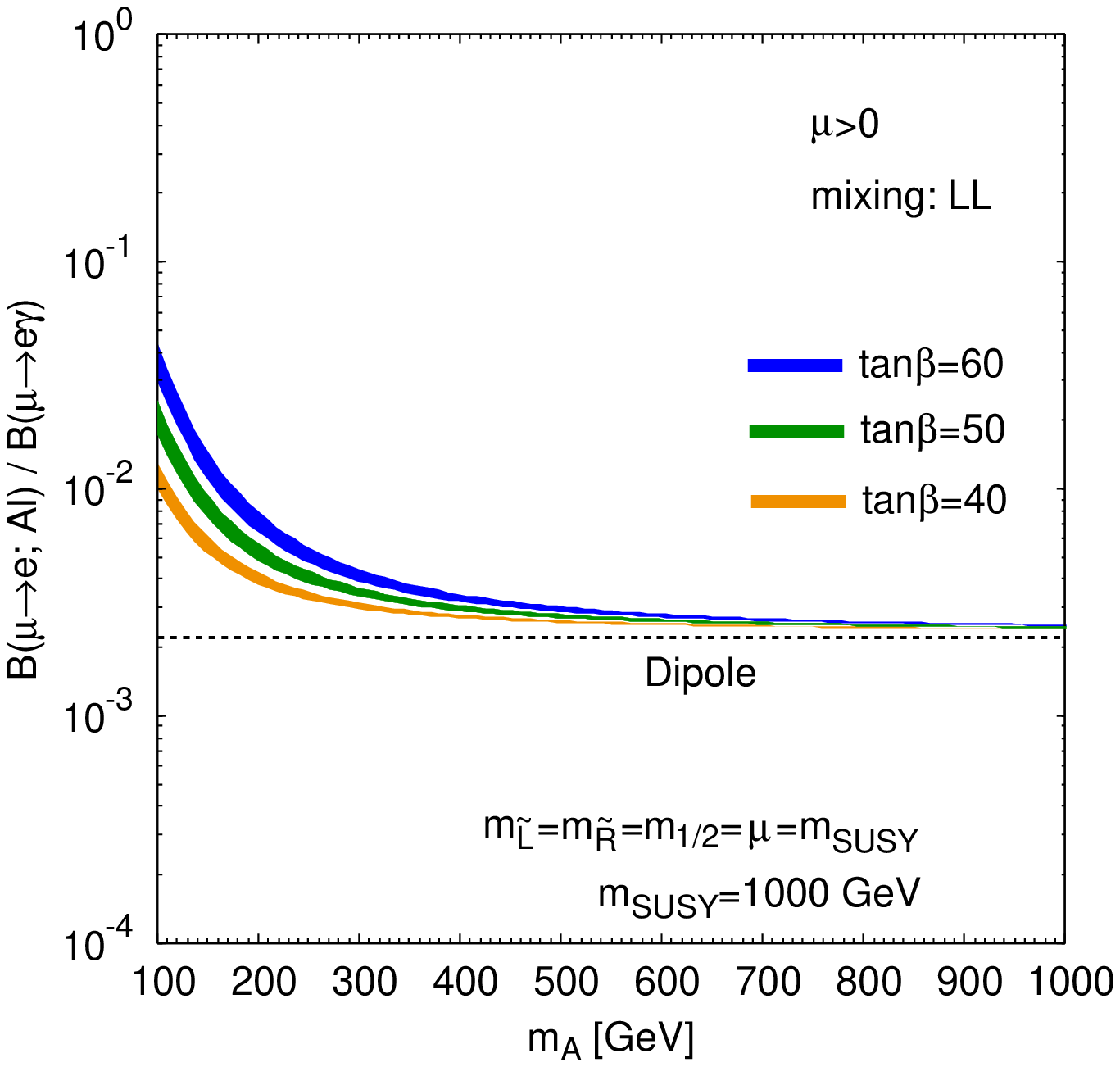}
\end{minipage}
\begin{minipage}{.49\linewidth}
\centering 
\includegraphics[width=7cm]{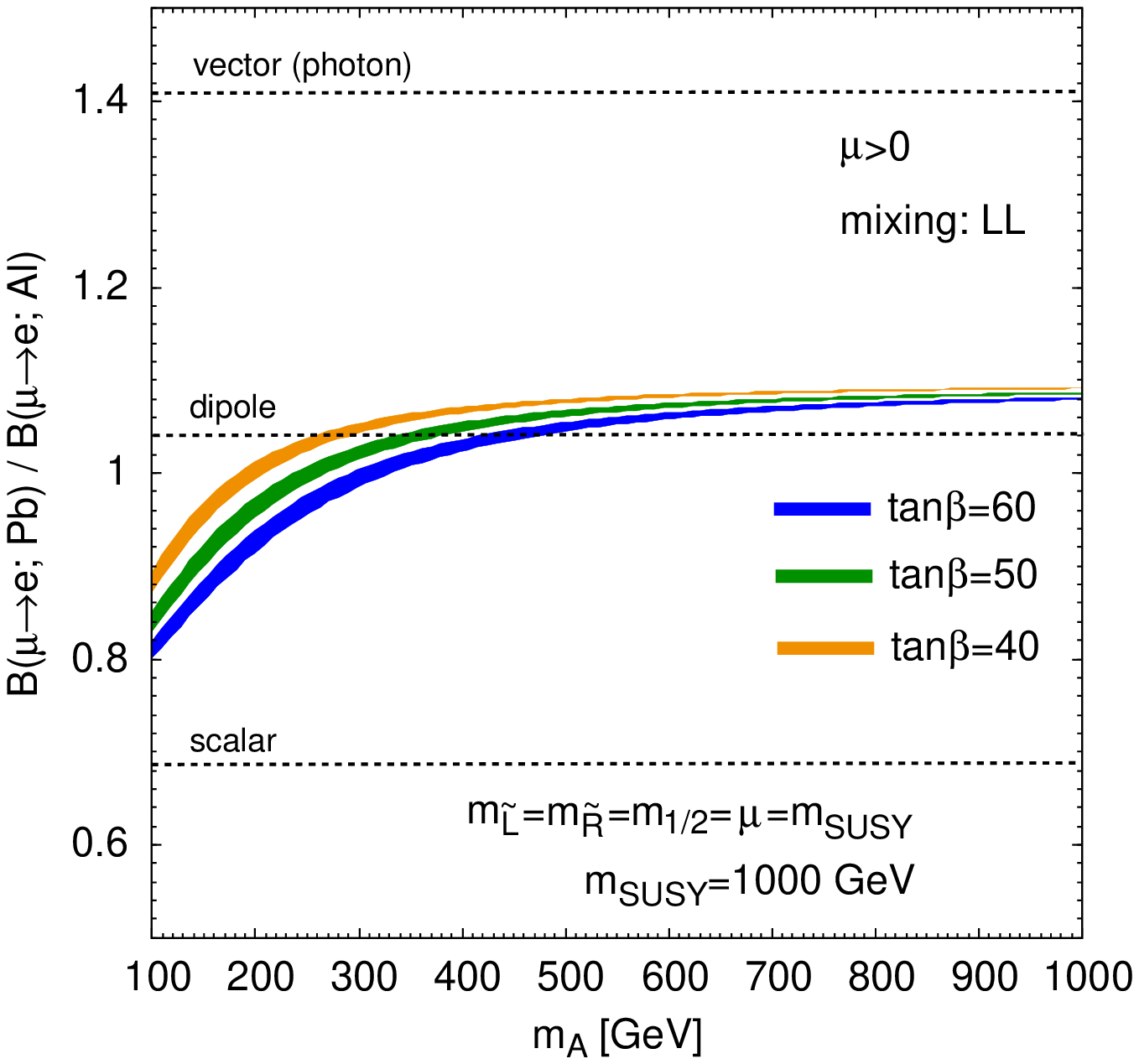}
\end{minipage}
\caption{ The pseudoscalar-Higgs mass dependence of 
$B_{\mu \to e} ({\rm Al}) / B_{\mu \to e \gamma} $ (left) 
and 
$B_{\mu \to e} ({\rm Pb}) / B_{\mu \to e} ({\rm Al})$  (right) 
for different values of $\tan \beta$.  The thickness of the bands 
corresponding to varying $y$ in the optimistic range $[0,0.05]$. } \label{fig:SUSY3}
\end{figure}

In both cases, the theoretical uncertainty becomes significant as the scalar operator
gets important.   In the context of this explicit supersymmetric model, 
a precise determination of the $y$ parameter is quite
important in order to extract information on  the underlying model parameters.          
To illustrate this even more explicitly,   in Fig.~\ref{fig:SUSY3}  
we show for $\mu > 0$ the $m_A$ 
dependence of  $B_{\mu \to e} ({\rm Al}) / B_{\mu \to e \gamma}$ 
(left)  and $B_{\mu \to e} ({\rm Pb}) / B_{\mu \to e} ({\rm Al})$ (right) 
for different values of $\tan \beta = 40, 50, 60$. 
In these plots, only the small uncertainty window is reported ($y \in [0,0.05]$), 
to illustrate the enhanced discriminating power.

\section{Conclusions}
\label{sect:conclusions}

In this work we have investigated whether the target-dependence of $\mu$-to-$e$ conversion 
rate  can be exploited to discriminate among underlying dynamical mechanisms of 
lepton flavor violation,  once one takes into account realistic hadronic and nuclear uncertainties.   
The major source of theoretical uncertainty  arises from  the 
nucleon matrix element of the strange quark scalar density. This is  expressed in terms of 
the parameter $y$ (see Eq.~\ref{eq:ypar}), which we have varied within 
two ranges reflecting the current uncertainty  ($y \in[0,0.4]$) and the projected 
uncertainty  within reach of lattice QCD calculations ($y \in [0,0.05]$).

In order to assess the model discriminating power of 
a combined phenomenological analysis of 
$\mu \to e \gamma$ and  $\mu \to e$ conversion on different target nuclei, 
we have defined  four classes of models,  
in which only one underlying short distance operator dominates over all the others
(Dipole, Scalar, Vector ($\gamma$) and Vector ($Z$)). 
Ratios of LFV branching fractions can be used to test the various models. 
The single-operator hypothesis  can be tested with at least one ratio (two LFV measurements), 
while the two-operator models, where two operators have comparable strength and interfere, 
can be tested with at least two ratios (three LFV measurements). 

Our conclusions are encouraging: the theoretical uncertainties 
(even at the current level)  are not an issue in 
testing the  single-operator dominance model,  
as they largely cancel when we take  ratios of  different conversion rates. 
On the other hand,  the current uncertainty  prevents meaningful tests of 
two-operator models involving  the Scalar operator,  
as it produces   errors of up to one order of magnitude 
in the double ratios in the interference region.
However,  with the anticipated reduced lattice error on the strange 
content of the nucleon,  this will not be an issue in the future. 
We have illustrated these main conclusions 
also  in the context of a suprsymmetric  model. 

Having established that the hadronic uncertainties will not be a limiting factor, 
we can ask how well one should measure the LFV rates in order 
to discriminate the underlying  models.  
Fig.~\ref{fig:target-dependence-1} shows that 
a realistic discrimination among single-operator  models requires a measure of 
the ratio of conversion rates in light nuclei (such as $B_{\mu \to e} ({\rm Ti})/B_{\mu \to e} ({\rm Al})$) 
at the level of $5 \%$ or better. 
Alternatively,  one would  need to  measure the ratio 
of conversion rates in a heavy and light element 
(such as  $B_{\mu \to e} ({\rm Pb})/B_{\mu \to e} ({\rm Al})$)  at the $20 \%$ level. 
Similar accuracy is required on the experimental side to be sensitive to 
interference effects when more than one operator is at work. 
Whether these challenging benchmark numbers can be reached in the 
future round of experiments~\cite{mu2e,comet} depends on many issues,  including the value of the 
branching fraction themselves (we are concerned here with ratios). 
Nonetheless,  we hope that our results will stimulate further experimental efforts towards   
measurements of $\mu \to e \gamma$ and $\mu$-to-$e$ conversion 
and consideration of various  options for target nuclei. 

\vspace{1.0cm}
{\bf Acknowledgments}  --  
We thank  Satoshi Mihara for useful exchanges on the 
experimental aspects of $\mu$-to-$e$ conversion 
and Tetsuya  Onogi  for useful comments on the sigma term. 
V.\ C.\  thanks the KEK theory group for its hospitality and support, 
as well as the Institute for Nuclear Theory at the University of 
Washington for its hospitality  and the Department of Energy for support during the 
completion of this  work. 
Y.\ O.\ is supported in part by the Grant-in-Aid for
Science Research, Ministry of Education, Culture, Sports, Science and
Technology, Japan, No.\ 16081211 and by the  Grant-in-Aid for Science
Research, Japan Society for the Promotion of Science, No.\ 20244037.
P.\ T.\   has been supported in part by the theory group at LANL (during her visit),
by a grant from the Spanish Ministry of Science and Innovation (AP2006-04522), 
as well as by the EU MRTN-CT-2006-035482  (FLAVIAnet),  by MEC (Spain) under grant FPA2007-60323, and by the  Spanish Consolider-Ingeino 2010 Programme CPAN (CSD2007-00042).


\begin{thebibliography}{99}


\bibitem{muegamma99} M. L. Brooks {\it et al.}
Phys. Rev. Lett. {\bf 83}, 1521 (1999).

\bibitem{mu3e88} U. Bellgardt {\it et al.}  Nucl. Phys. {\bf B299}, 1  (1988).
  
\bibitem{mueconvAu} W. Bertl  {\it et al.}
PSI annual report, p.9 (2002) (unpublished).

\bibitem{mueconvTi} C. Dohmen  {\it et al.}
Phys. Lett. {\bf B317}, 631 (1993).

\bibitem{mueconvPb} W. Honecker {\it et al.}
Phys. Rev. Lett. {\bf 76}, 200 (1996).


\bibitem{ryuichiro} 
  R.~Kitano, M.~Koike and Y.~Okada,
  Phys.\ Rev.\  D {\bf 66}, 096002 (2002)
  [Erratum-ibid.\  D {\bf 76}, 059902 (2007)]
  [arXiv:hep-ph/0203110].

\bibitem{shanker}
  O.~U.~Shanker,
  Phys.\ Rev.\  D {\bf 20}, 1608 (1979).
  
\bibitem{marciano}  
  A.~Czarnecki, W.~J.~Marciano and K.~Melnikov,
  AIP Conf.\ Proc.\  {\bf 435}, 409 (1998)
  [arXiv:hep-ph/9801218].



\bibitem{Kosmas:2001mv}
  T.~S.~Kosmas, S.~Kovalenko and I.~Schmidt,
  Phys.\ Lett.\  B {\bf 511}, 203 (2001)
  [arXiv:hep-ph/0102101].


\bibitem{Shifman:1978zn}
  M.~A.~Shifman, A.~I.~Vainshtein and V.~I.~Zakharov,
  Phys.\ Lett.\  B {\bf 78}, 443 (1978).


	
\bibitem{cors-nath} 
  A.~Corsetti and P.~Nath,
  Phys.\ Rev.\  D {\bf 64}, 125010 (2001)
  [arXiv:hep-ph/0003186].




\bibitem{onogi} 
  H.~Ohki {\it et al.},
  Phys.\ Rev.\  D {\bf 78}, 054502 (2008)
  [arXiv:0806.4744 [hep-lat]].


	
	
\bibitem{gls}
  J.~Gasser, H.~Leutwyler, M.~P.~Locher and M.~E.~Sainio,
  Phys.\ Lett.\  B {\bf 213}, 85 (1988); 
  J.~Gasser, H.~Leutwyler and M.~E.~Sainio,
  Phys.\ Lett.\  B {\bf 253}, 252 (1991).

\bibitem{pavan} 
  M.~M.~Pavan, I.~I.~Strakovsky, R.~L.~Workman and R.~A.~Arndt,
  PiN Newslett.\  {\bf 16}, 110 (2002)
  [arXiv:hep-ph/0111066].

	


\bibitem{borasoy} 
  B.~Borasoy and U.~G.~Meissner,
  Annals Phys.\  {\bf 254}, 192 (1997)
  [arXiv:hep-ph/9607432].


\bibitem{procura04}  
  M.~Procura, T.~R.~Hemmert and W.~Weise,
  Phys.\ Rev.\  D {\bf 69}, 034505 (2004)
  [arXiv:hep-lat/0309020]; 
%
  M.~Procura, B.~U.~Musch, T.~Wollenweber, T.~R.~Hemmert and W.~Weise,
  Phys.\ Rev.\  D {\bf 73}, 114510 (2006)
  [arXiv:hep-lat/0603001].


\bibitem{cheng89}
H. Cheng, Phys.\ Lett. \ {\bf B219}, 347 (1989).

\bibitem{leutwyler}
  H.~Leutwyler,
  Phys.\ Lett.\  B {\bf 378}, 313 (1996)
  [arXiv:hep-ph/9602366].


\bibitem{GUT1}
  R.~Barbieri and L.~J.~Hall,
  Phys.\ Lett.\  B {\bf 338}, 212 (1994)
  [arXiv:hep-ph/9408406]; 
  R.~Barbieri, L.~J.~Hall and A.~Strumia,
  Nucl.\ Phys.\  B {\bf 445}, 219 (1995)
  [arXiv:hep-ph/9501334];  
%
  J.~Hisano, T.~Moroi, K.~Tobe and M.~Yamaguchi,
  Phys.\ Lett.\  B {\bf 391}, 341 (1997)
  [Erratum-ibid.\  B {\bf 397}, 357 (1997)]
  [arXiv:hep-ph/9605296].


\bibitem{seesaw1}
  F.~Borzumati and A.~Masiero,
  Phys.\ Rev.\ Lett.\  {\bf 57}, 961 (1986); 
  J.~Hisano, T.~Moroi, K.~Tobe and M.~Yamaguchi,
  Phys.\ Rev.\  D {\bf 53}, 2442 (1996)
  [arXiv:hep-ph/9510309].


\bibitem{Cirigliano:2004mv}
  V.~Cirigliano, A.~Kurylov, M.~J.~Ramsey-Musolf and P.~Vogel,
  Phys.\ Rev.\  D {\bf 70}, 075007 (2004)
  [arXiv:hep-ph/0404233].


\bibitem{largetanbeta1} 
  R.~Kitano, M.~Koike, S.~Komine and Y.~Okada,
  Phys.\ Lett.\  B {\bf 575}, 300 (2003)
  [arXiv:hep-ph/0308021].


\bibitem{RPVSUSY1}
  J.~E.~Kim, P.~Ko and D.~G.~Lee,
  Phys.\ Rev.\  D {\bf 56}, 100 (1997)
  [arXiv:hep-ph/9701381];  
%
  K.~Huitu, J.~Maalampi, M.~Raidal and A.~Santamaria,
  Phys.\ Lett.\  B {\bf 430}, 355 (1998)
  [arXiv:hep-ph/9712249];  
%
  A.~Faessler, T.~S.~Kosmas, S.~Kovalenko and J.~D.~Vergados,
  Nucl.\ Phys.\  B {\bf 587}, 25 (2000); 
%
  A.~de Gouvea, S.~Lola and K.~Tobe,
  Phys.\ Rev.\  D {\bf 63}, 035004 (2001)
  [arXiv:hep-ph/0008085]. 


\bibitem{dejager}
  H.~De Vries, C.~W.~De Jager and C.~De Vries,
  Atom.\ Data Nucl.\ Data Tabl.\  {\bf 36}, 495 (1987); 
G.~Fricke {\it et al.}, {\it ibid.} {\bf 60}, 177 (1995). 


\bibitem{mu2e}   
Mu2e Collaboration website: http://mu2e.fnal.gov/

\bibitem{comet} 
COMET:  http://comet.phys.sci.osaka-u.ac.jp/internal/publications/main.pdf

\end{thebibliography}
\end{document}